\documentclass[preprint]{emulateapj} 

\usepackage{natbib}
\usepackage{hyperref}
\hypersetup{colorlinks,citecolor=Blue,linkcolor=Red,urlcolor=Blue}
\usepackage{amsmath}
\usepackage{empheq}
\usepackage[usenames,dvipsnames]{color}
\DeclareMathOperator\erf{erf}
\allowdisplaybreaks 
\hypersetup{colorlinks,citecolor=Blue,linkcolor=Red,urlcolor=Blue}
\usepackage[usenames,dvipsnames]{color}
\usepackage{amsmath,amsfonts,amssymb}
\usepackage{graphicx}
\usepackage{wasysym}
\usepackage{float}
\usepackage{url}
\newcommand{\Poincare}{{Poincar$\acute{\rm{e}}$}}
\newcommand{\G}{\mathcal{G}}
\newcommand{\Ham}{\mathcal{H}}
\newcommand{\Hamp}{\mathcal{H'}}
\newcommand{\Hampb}{\bar{\mathcal{H}}'}
\newcommand{\appropto}{\mathrel{\vcenter{\offinterlineskip\halign{\hfil$##$\cr\propto\cr\noalign{\kern2pt}\sim\cr\noalign{\kern-2pt}}}}}

\begin{document}


\title{In Situ Formation and Dynamical Evolution of Hot Jupiter Systems}  
\author{Konstantin Batygin$^{1}$, Peter H. Bodenheimer$^{2}$, Gregory P. Laughlin$^{2}$} 

\affil{$^{1}$Division of Geological and Planetary Sciences, California Institute of Technology, Pasadena, CA 91125} 
\affil{$^{2}$UCO/Lick Observatory, University of California, Santa Cruz, California 95064} 
\email{kbatygin@gps.caltech.edu}

\begin{abstract} 
Hot Jupiters, giant extrasolar planets with orbital periods shorter than $\sim10\,$days, have long been thought to form at large radial distances, only to subsequently experience long-range inward migration. Here, we offer the contrasting view that a substantial fraction of the hot Jupiter population formed \textit{in situ} via the core accretion process. We show that under conditions appropriate to the inner regions of protoplanetary disks, rapid gas accretion can be initiated by Super-Earth type planets, comprising $10-20\,$Earth masses of refractory material. An \textit{in situ} formation scenario leads to testable consequences, including the expectation that hot Jupiters should frequently be accompanied by additional low-mass planets with periods shorter than $\sim100\,$days. Our calculations further demonstrate that dynamical interactions during the early stages of planetary systems' lifetimes should increase the inclinations of such companions, rendering transits rare. High-precision radial velocity monitoring provides the best prospect for their detection.
\end{abstract} 

\maketitle
\section{Introduction}
The 1995 discovery of 51 Peg\,b -- a 0.5 $M_{\rm J}$ planet on a $P=4.5\,$day orbit around a nearby solar-type star \citep{MayorQueloz1995} was a genuine surprise. Prior to the detection of extrasolar planets, theories of planet formation were almost solely informed by the architecture of the present-day solar system. In particular, it was expected that giant planet formation occurs beyond the nebular snow line, where icy material is available to form multi-Earth mass protoplanetary cores that subsequently experience rapid gas accretion \citep{Pollacketal1996}. 

The existence of hot Jupiters was attributed, starting with \citet{Linetal1996}, to giant planet conglomeration at large distances ($a\gtrsim 2-5\,$AU), followed by extensive inward migration. Over the past two decades, this sequence of events has become an established theoretical paradigm, and considerable effort has been dedicated to determining the precise nature of the dominant mode of orbital transport \citep{WuMurray2003,KleyNelson2012,BeaugeNesvorny2012}. Here, we consider the possibility that many hot Jupiter-class planets actually form \textit{in situ}, via the core accretion mechanism at close-in orbits, and outline a set of observational tests that can falsify our hypothesis.

\begin{figure*}[t]
\centering
\includegraphics[width=0.7\textwidth]{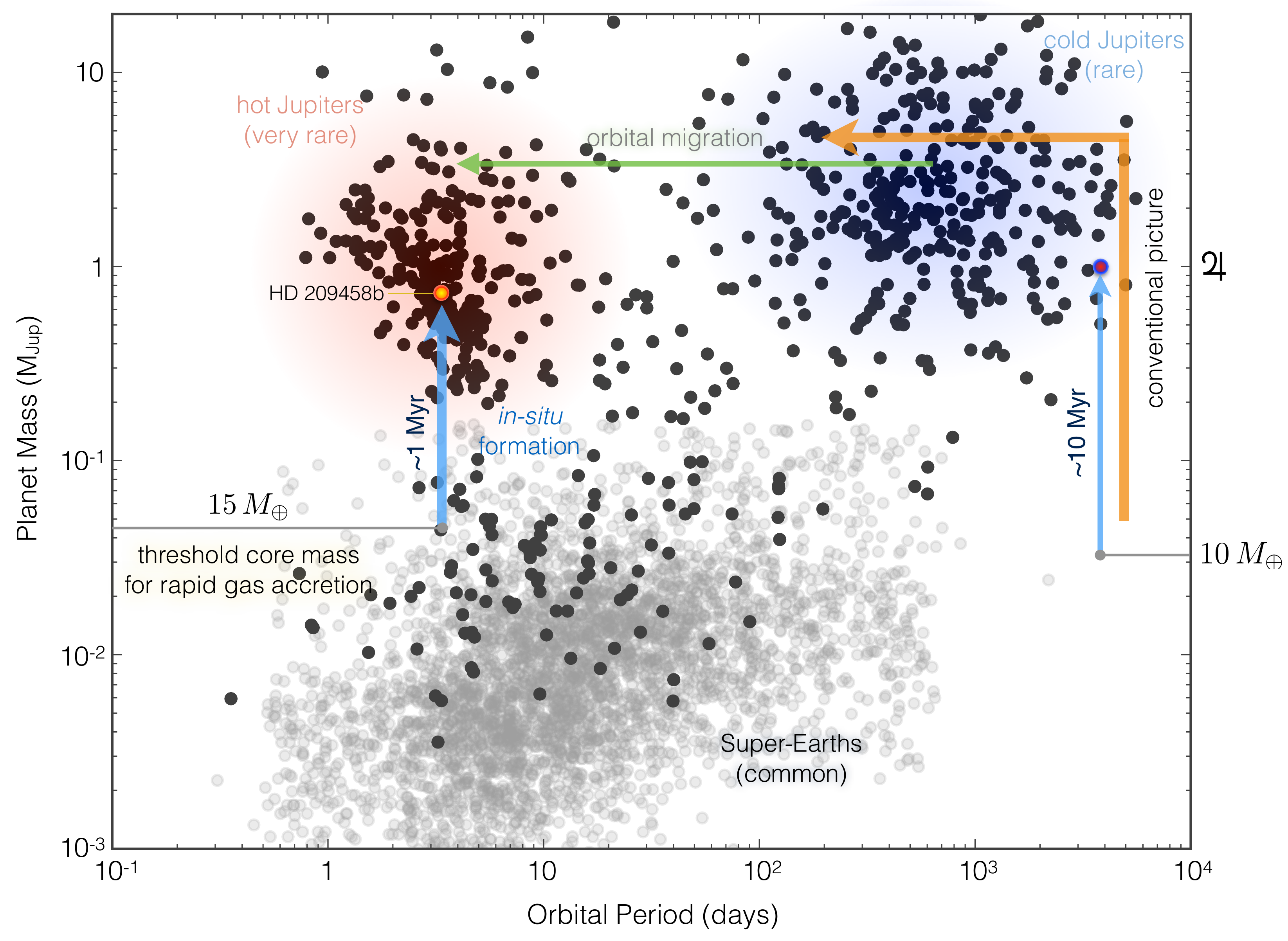}
\caption{A schematic mass-period diagram that represents the current extrasolar planetary catalog. Objects with directly measured masses ($m\,\sin(i)$ for non-transiting planets) are shown as solids points, while objects with masses that are inferred indirectly from radius measurements are depicted as transparent dots. The conventional narrative for hot Jupiter formation affirms that the observed close-in giant planets initially form via core-nucleated accretion as members of the cold Jupiter population, and subsequently migrate inward via interactions with the protoplentary disk or dynamic excitation. Our calculations suggest that hot Jupiter conglomeration can instead proceed \textit{in situ}, on a timescale of order $\sim 1\,$Myr $-$ considerably shorter than the typical lifetimes of protoplanetary disks. Accordingly, within the framework of our model, close-in Super-Earths comprise the source population of hot Jupiter cores, and the threshold planet mass needed to trigger runaway accretion at an orbital semi-major axis of $a\sim0.05\,$AU is approximately $\sim15\,M_{\oplus}$.}
\label{fig:MRfig}
\end{figure*}

Our motivation to consider \textit{in situ} formation as an alternative to long-range migration is two-fold. First, if hot Jupiters are derived from a population of distant, Jupiter-like planets, the physical properties of hot and cold giant planets should, on average be the same. Figure (\ref{fig:MRfig}) shows the current exoplanet census, with contributions from Doppler velocity surveys (in which minimum masses, $m \sin(i)$, are determined directly), as well as from transit timing and transit-based radius measurements. There is a pronounced concentration of hot Jupiter class planets with orbital periods close to three days and masses slightly \textit{below} that of Jupiter. Simultaneously, the average giant planet with period $100\,{\rm d}<P<3000\,{\rm d}$ (a population which is associated with $\sim$5-10\% of solar-type stars in the Sun's vicinity) is several times \textit{more massive} than Jupiter. Even after accounting for observational bias that favors detection of more massive bodies at large radii, this disparity suggests that close-in and distant giant planet populations are intrinsically distinct \citep{Knutsonetal2014,Bryan2016}. 

This division appears to be inconsistent with smooth disk-driven migration, since the rate of orbital decay for giant, gap-opening planets is (to leading order) independent of the planetary mass \citep{KleyNelson2012}. In some contrast, a subset of mechanisms associated with the high-eccentricity channel of orbital migration (e.g. the secular chaos model; \citealt{WuLith2011}) do preferentially produce hot Jupiters with lower masses. However, the analysis of \citet{Dawson2015} convincingly demonstrates that the high-eccentricity channel is generally disfavored as a dominant route for hot Jupiter generation. Therefore, one can speculate that the observed difference in the average masses of the hot and cold Jupiter populations stems not from migration, but from varying conglomeration sites.

A second issue of relevance is related to the primordial mass-budget of protoplanetary material. Traditionally, \textit{in situ} conglomeration has been discounted because of a lack of sufficiently massive cores at small orbital radii \citep{Rafikov2006}. This concern is rooted in the so-called Minimum Mass Solar Nebula \citep{Hayashi1981}, in which the primordial solar system contains essentially no solid material between the Sun and Venus. However, Doppler velocity surveys \citep{Howard2010,Mayoretal2011} and space-based transit photometry \citep{Batalhaetal2013} have demonstrated that inner regions of planetary systems are not generally empty. Instead, hot ``Super-Earths,'' planets with $P<100\,$days and $m<30M_{\oplus}$ are extremely common, with 30-50\% of main sequence stars in the solar neighborhood serving as hosts \citep{Fressin2013,DressingCharbonneau2013,DressingCharbonneau2015,Petigura2013,Mulders2015,Winn2015}. 

It is worth noting that Super-Earths frequently occur in multi-planet configurations, with member planets exhibiting small mutual inclinations, small eccentricities, and a relative dearth of low-order mean-motion resonances \citep{Lissauer2012,FangMargot2012,Fabryckyetal2014,Rowe2014}. The overwhelming prevalence of such systems suggests that they represent the default outcome of planet formation. The seeming discrepancy between such conditions and those inferred from the solar nebula can be resolved by accounting for Jupiter's inward-then-outward ``Grand Tack'' \citep{Walshetal2011}, which may have precipitated the destruction of any early-forming Super-Earths within our own system \citep{BatyginLaughlin2015}.

Early analyses of the core-accretion mechanism demonstrated that the critical core mass required to trigger runaway accretion only exhibits weak\footnote{Under the assumption of a purely radiative envelope, this dependence is logarithmic. More realistic envelope models, possessing both radiative and convective layers, are characterized by a weak power-law dependence.} dependence on the surrounding nebular conditions \citep{Stevenson1982,Rafikov2006}. The masses of protoplanetary cores required to trigger runaway gas accretion at the high ambient temperatures (such as those associated with hot Jupiter orbits) can thus be expected to be similar to threshold core masses needed to initiate the formation of Jupiter or Saturn \citep{Bodenheimer2000, LeeChiang2014}. 

A separate question concerns the availability of gaseous material. Some efforts aimed at reconstruction of the minimum mass extrasolar nebula \citep{ChiangLaughlin2013,Schlichting2014} allow for sufficient mass to exist in the inner disk for isolated accretion. Moreover, even if the nebular mass budget is locally insufficient to form a bonafide giant planet at $a\sim0.05\,$ AU, gaseous material is continuously resupplied to close-in orbits over the lifetime of the disk by viscous accretion.

With these ideas in mind, we can examine the physical process of \textit{in situ} hot Jupiter formation and its dynamical consequences quantitatively. The paper is structured as follows. In section 2, we model the accretion of gas onto protoplanetary cores under nebular conditions appropriate to a stellocentric distance of typical hot Jupiter orbits. In section 3, we present a dynamical study of the behavior of Super-Earth type planets (modeled as test particles) within the potential generated by a rapidly rotating young star and a proto-hot Jupiter that experiences \textit{in situ} growth. We summarize and discuss the implications of our results in section 4.

\section{\textit{In Situ} Conglomeration of Hot Jupiters}

In order to assess the viability of the \textit{in situ} formation scenario for hot Jupiters, we have carried out a series of numerical calculations of the core accretion process under conditions characteristic of the high-density, high-temperature inner region of the protostellar disk. Our models employ the Henyey technique \citep{Henyey1964} to simulate the thermal evolution and mass accretion onto protoplanetary gaseous envelopes, using the standard equations of stellar structure, and the assumption of a spherically symmetric growing planet. Our code has been extensively discussed in the literature, and has been used to calculate the formation phases of both, solar system planets (e.g. \citealt{Pollacketal1996,Hubickyj2005,Helled2014}), as well as extrasolar planets \citep{Bodenheimer2003,Dodson-Robinson2009,Rogers2011}.

Following the work of \citet{Bodenheimer2000}, where in-situ formation of hot Jupiters was first proposed, a number of studies have considered core-nucleated accretion of giant planets in close proximity to the host star (see e.g. \citealt{Ikoma2001,Boley2016}). Recently, \citet{LeeChiang2014} considered runaway accretion initiated by close-in Super-Earths, and found that 10 $M_{\oplus}$ planets can initiate runaway accretion at $a\sim 0.1\,$AU on $\sim$ Myr timescales\footnote{We note that the calculations of \citet{LeeChiang2014,LeeChiang2015} are subject to a different set of physical assumptions than the calculations presented in this work.}. In follow-up studies, \citet{LeeChiang2015,LeeChiang2016} found that critical core masses for runaway accretion can range between $2-8$ $M_{\oplus}$, depending on envelope metallicities. Consequently, \citet{LeeChiang2016} proposed that the vast majority of Super-Earths must form in transitional disks, after the gas has been largely depleted. 

We have carried out four evolutionary simulations that span a similar range of core masses. Specifically, we start with solid protoplanetary cores comprising 4 $M_{\oplus}$ (case 1), 10 $M_{\oplus}$ (case 2), and 15 $M_{\oplus}$ (cases 3 and 4) to seed giant planet conglomeration. The timescale for core formation in each model was taken to be $200\,$kyr, while the semi-major axis of the planetary orbit was fixed at $a=0.05\,{\rm AU}$. We purposely avoid specifying the physical formation mechanism of the solid core itself. The origins of the Super-Earth population are widely debated, and both migration-based models \citep{TerquemPapaloizou2007,CresswellNelson2008} as well as \textit{in situ} accretion scenarios \citep{HansenMurray2012,ChiangLaughlin2013} have been proposed to explain the data. Importantly, our results are largely independent of how the solid core arises, as long as the formation timescale does not exceed the lifetime of the disk \citep{LeeChiang2015}. Rapid conversion of planetesimals into Super-Earths by accretion of pebbles appears to be fully consistent with the envisioned evolutionary sequence \citep{LambrechtsJohansen2012,LambrechtsJohansen2014,Levison2015}.

For definitiveness, we assume a central mass $M_{\star}=1\,M_{\odot}$ and adopt a nebular temperature of 1500K (for full details of the core growth calculation and physical assumptions, see \citealt{Hubickyj2005}). The outer radius of the growing planet was taken to be the Hill radius\footnote{Strictly speaking, the assumption of spherical symmetry inherent to the employed model is only well justified as long as the Hill radius does not exceed the disk scale height (i.e. $r_H/a<h/a\sim0.05-0.1$; \citealt{Armitage2011}). Quantitatively, this corresponds to planet masses that are already in the giant planet regime, so this point is not crucial to the question of whether or not runaway gas accretion can be triggered on the appropriate timescale.}, 
\begin{align}
r_H=a \left(\frac{m}{3M_{\star}}\right)^{1/3}.
\end{align}
Dust opacities in the protoplanetary envelopes were assumed to be interstellar (e.g. \citealt{Podolak2003}), although they have little importance in this context, due to the high ambient temperature. The equation of state presented by \citet{Saumon1995} was employed for the gaseous envelope.

\begin{figure}
\includegraphics[width=\columnwidth]{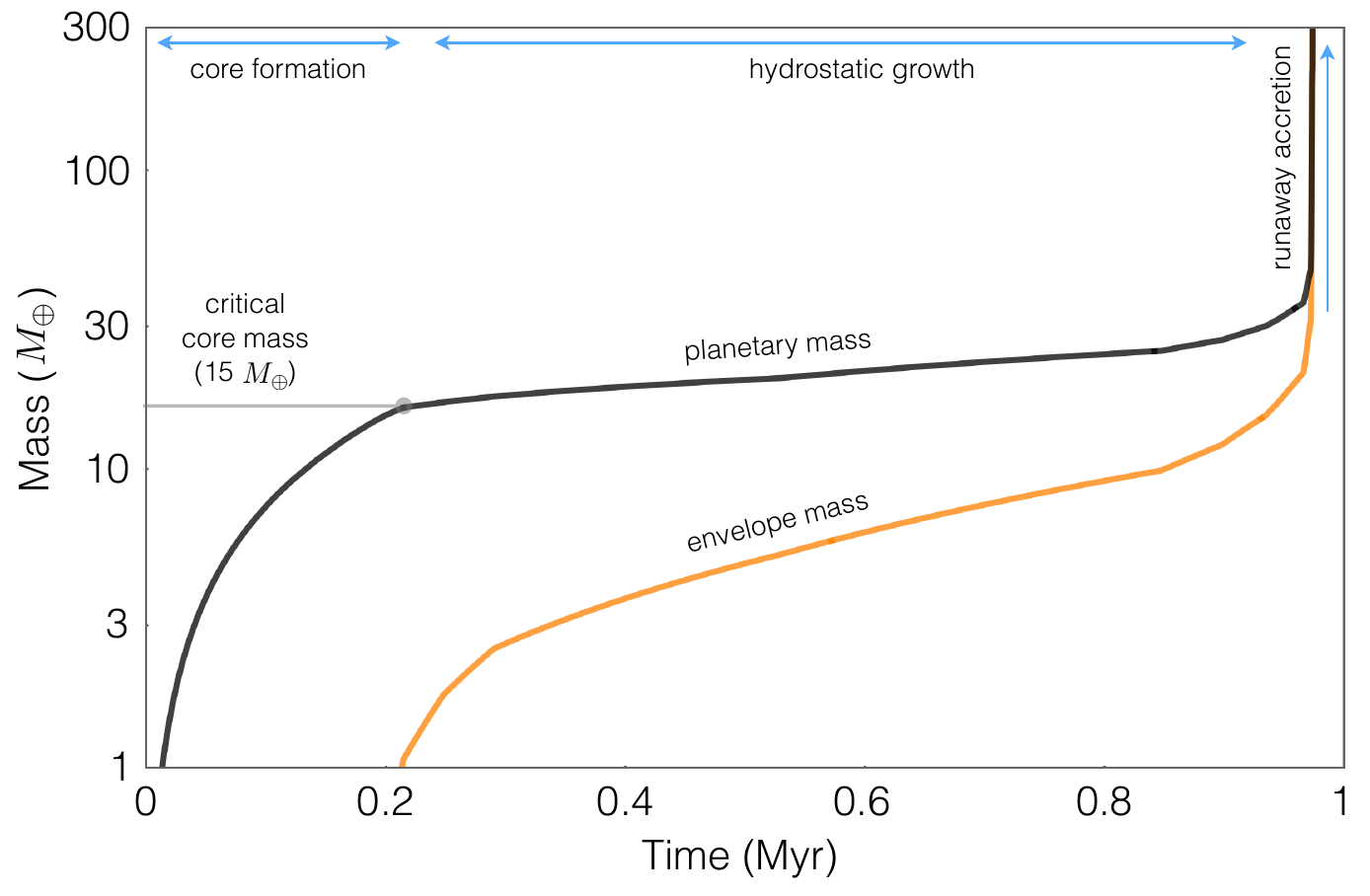}
\caption{Evolutionary track associated with \textit{in situ} conglomeration of a hot Jupiter at a close-in orbital radius. The orange curve depicts the mass of the gaseous envelope, while the black curve represents the total planetary mass. Following the formation of a refractory composition $15\,M_{\oplus}$ core, the body experiences hydrostatic growth for approximately $0.8\,$Myr. As the envelope mass reaches a value comparable to that of the core, an episode of runaway accretion is triggered, rapidly producing a highly irradiated giant planet.}
\label{pollackfig}
\end{figure}

Figure (\ref{pollackfig}) shows the results corresponding to case 3. Following an initial phase of solid body accretion, the growing core is immersed in a nebular disk with density of $\rho=1.37\times10^{-6}\,$g$\,$cm$^{-3}$, equivalent to a gas disk surface density of $\Sigma_{\rm{gas}}\sim10^{5}\,$g$\,$cm$^{-2}$ and a solid disk surface density of order $\Sigma_{\rm{solid}}\sim10^{3}\,$g$\,$cm$^{-2}$. Correspondingly, the ambient nebular surface density in our model is roughly a factor of two lower than that predicted by the inward extrapolation of the minimum mass solar nebula \citep{Hayashi1981}, which yields $\rho_{\rm MMSN}\sim3\times10^{-6}\,$g$\,$cm$^{-3}$.

During the period ranging from $200\,$kyr to $\sim900\,$kyr, the mass of the planet's gaseous envelope gradually increases, with a growth rate that is limited by its Kelvin-Helmholtz contraction. At time $t=934\,$kyr, the mass of the envelope reaches a value equal to the mass of the core, i.e. 15$\,M_{\oplus}$. The occurrence of this ``crossover,'' in which the envelope mass exceeds the core mass, ushers in a period of rapid gas accretion. Within $50\,$kyr, the planet reaches a typical hot Jupiter mass of 0.7$M_{\rm Jup}$.

In case 1, we find that a 4 $M_{\oplus}$ core, which is subject to the same physical assumptions as case 3, experiences very little gas accretion. After a total evolution time of $3\,$Myr, the envelope consists of only 0.19$M_{\oplus}$. Case 2, in which the solid accretion phase generates a 10 $M_{\oplus}$ core, achieves an envelope mass of 3.14 $M_{\oplus}$ after $3\,$Myr. While substantial, this gas mass is insufficient to trigger runaway growth of the planet to hot Jupiter proportions. Finally, in case 4, the nebular density was lowered to $\rho=4.4\times10^{-7}\,$g$\,$cm$^{-3}$. Despite this lower ambient gas density, the planet entered a phase of runaway gas accretion after a total time of $1.4\,$Myr, indicating that the nebular gas density plays a muted role in determining whether a forming planet can become a hot Jupiter via \textit{in situ} accretion.

Although we have not restricted the local mass supply of the nebula in our simulations, it is important to note that in reality, the disk material must be delivered to the growing protoplanet from more distant orbits by viscous accretion. This implies that the derived conglomeration timescales should be viewed as lower bounds. Yet, because gas accretion is exceedingly rapid once a planet achieves the runaway phase, it is very likely that typical protoplanetary disks can resupply the required material at a sufficiently rapid rate, as to not alter our estimates dramatically. 

We note that there exist additional physical mechanisms that can contribute to the sculpting of the final mass-period diagram, that our calculations do not capture. For example, envelope heating by planetesimal accretion could significantly prolong the phase of hydrostatic growth \citep{Ikoma2000}, rendering our estimate of the critical core mass an effective lower bound. Moreover, the theory that we outline here does not provide an immediate explanation for a slope in the upper boundary traced by hot Jupiters in the mass period plane, as discussed in detail by \citet{Mazeh2016}. If this sloped boundary is a robust feature of the volume-limited planet distribution, then it could constitute a point of evidence against our model in its current form. Simultaneously, processes such as photo-evaporation and magnetospheric coupling can act to carve out the so-called ÒSuper-Earth desertÓ at very short orbital periods (see \citealt{Adams2011,Lundkvist2016} and the references therein). Correspondingly, a full account of such effects is required to understand the origins of the extrasolar mass-period distribution at a detailed level. 

Because of their ready detectability, hot Jupiters are strongly overrepresented in the current planetary census. Volume-limited surveys indicate that their true occurrence rate is of order $0.5-1$\% \citep{UdrySantos2007}. This infrequency is consistent with the above picture where only the most massive short-period Super-Earths are able to initiate runaway accretion. We note, however, that this interpretation is not unique. For instance, the paucity of close-in giant planets can be alternatively be explained by the relative timing of gas dispersal and core coagulation, as advocated by \citet{LeeChiang2014,LeeChiang2016}. 

Undeniably, the nebular abundance of heavy elements also plays a role. Hot Jupiters show a strong increase in occurrence fraction as a function of host-star metallicity. Therefore, our the proposed \textit{in-situ} formation scenario insinuates that a mass-metallicity correlation should also exist among Super-Earths with periods $1\,{\rm d}<P<10\,{\rm d}$. Indeed, statistical evidence has been marshaled in support of the existence of a positive correlation between super Earth radii (and by extension, planetary mass) and stellar metallicity \citep{Buchhave2014,WangFischer2015}. Recent detailed work by \citet{DawsonChiangLee2015} discusses how such correlations can be understood; either via \textit{in situ} mechanisms, or through planet formation frameworks that include migration.

Due to the innate multiplicity of hot low-mass planets, \textit{in situ} formation implies that many hot Jupiters experience conglomeration within systems that also contain Super-Earths. A critical question, then, concerns the dynamical fate of these lower-mass companions. Do they exist? If so, what are their properties and prospects for detection? Conventional wisdom \citep{Steffenetal2012} holds that hot Jupiters are rarely accompanied by additional objects (low-mass or otherwise) having orbital periods $P\lesssim100$ days. As a consequence, the identification of unique, unexpected properties of prospective companion planets to hot Jupiters would constitute a useful test of the \textit{in situ} formation scenario.

\section{Dynamical Evolution of Compact Planetary Systems}

Having demonstrated the physical viability of $\textit{in situ}$ accretion of hot Jupiters, we now wish to understand how the conglomeration process affects nearby planets. Our focus lies in describing the dynamical evolution inherent to the concluding stages of planet formation, and the discussion is tailored towards \textit{in situ} growth of giant planets. Accordingly, the effects that will be of dominant importance are limited to accretion of planetary mass \citep{Pollacketal1996} as well as rotational evolution and Kelvin-Helmholtz contraction of the young star \citep{Bouvier2013}. We initially ignore the auxiliary effects arising from the presence of a protoplanetary disk (see e.g. \citealt{PapaloizouLarwood2000}). However, as we argue below, the consequences of disk-driven evolution are by no means central to the presented ideas, and only affect the results on a detailed level.

\subsection{Analytical Theory}

\subsubsection{Inclination Evolution}

\begin{figure}[t]
\centering
\includegraphics[width=\columnwidth]{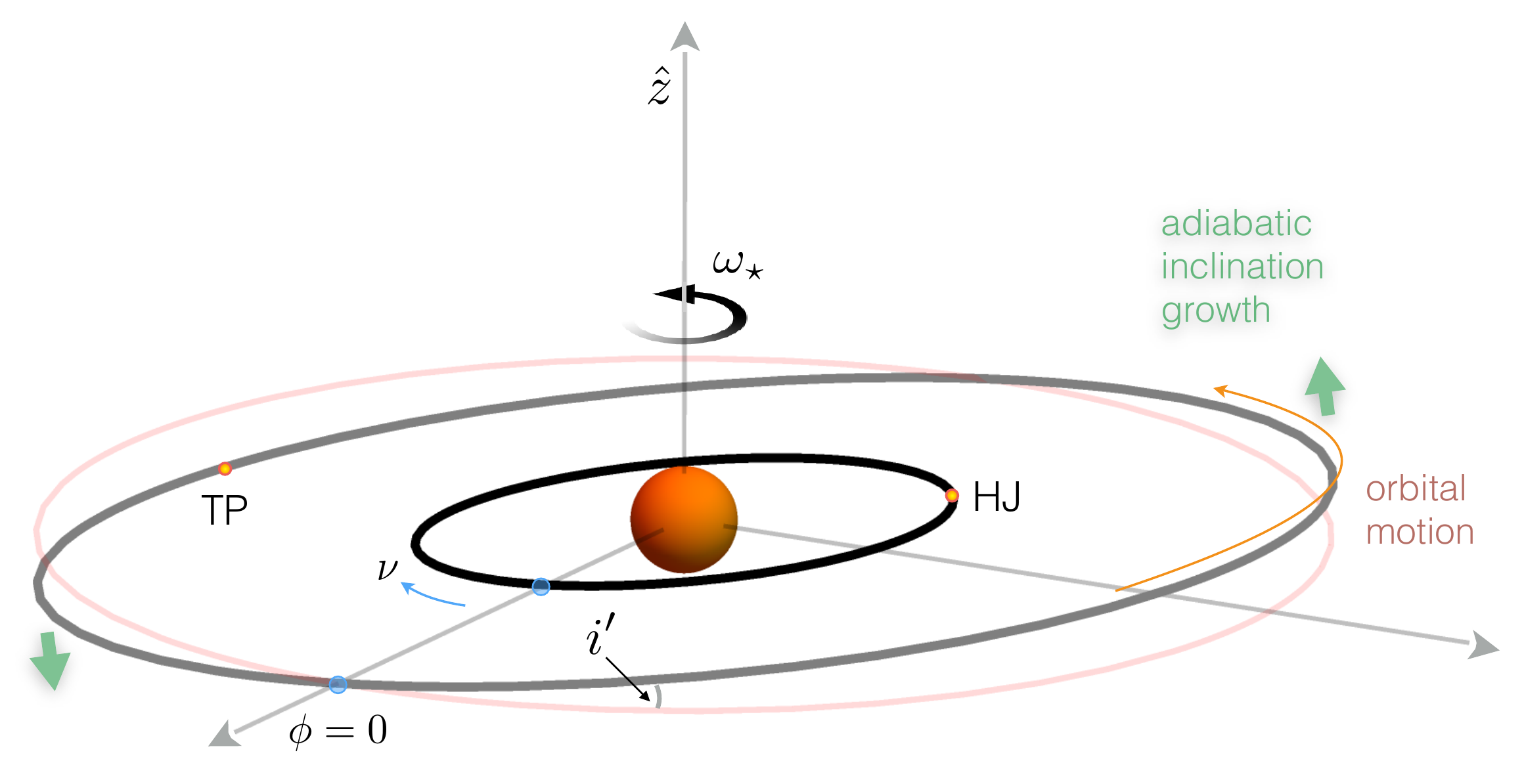}
\caption{A cartoon depicting the geometrical setup of the analytical calculation. A $R_{\star} = 2 R_{\odot}$ star, orbited by a massive planet on a $a=0.05$ AU orbit, as well as a test particle on a $a'=0.12$AU orbit are shown, to scale. The coordinate system is chosen such that the $\hat{z}$-axis coincides with the (fixed) spin axis of the star, and the $\hat{x}$-axis regresses with the node of the inner planet at a rate $\nu$ (see equations \ref{nu} and \ref{transfomr}). }
\label{setupfig}
\end{figure}

Observations demonstrate that the orbital structure of multi-planet systems generally exhibits great diversity \citep{Mayoretal2011,Batalhaetal2013}. Here, we make no attempt at a population synthesis-type study (see e.g. \citealt{IdaLin2005}), and instead aim to elucidate a single, fundamental process that acts to sculpt the final architectures of the observed planetary systems. To this end, it seems sensible to explore the simplest non-trivial realization of a planetary system that exhibits the desired behavior (that is, the spatial, secular, circular, restricted, oblate three body problem - see Figure \ref{setupfig}). We begin within the framework of perturbation theory, and then progress to numerical experiments.

Throughout the manuscript, we adopt the following notation for similarly named orbital variables (e.g. $a, m, i,$ etc.). Variables corresponding to the central star will be labeled with a $\star$ subscript. Variables that represent the stellar angular momentum vector will be labeled with a tilde. Unmarked variables will correspond to a massive planet. Finally, variables related to the test particle will be denoted with a prime.

Consider the dynamical evolution of a pair of low-mass planets, orbiting a rapidly rotating, oblate T-Tauri star. Hot Jupiters are routinely found at orbital radii similar to the inferred truncation radii of protoplanetary disks. Correspondingly, we consider a system where only the inner-most planet reaches the runaway accretion phase within the nebular lifetime.

For definitiveness, we take the initial masses of the planets to be null, while allowing only the inner planet to accrete mass in time (i.e. the outer object remains a test-particle). Moreover, we conform to a hierarchy in the primordial angular momentum budget wherein the stellar spin holds the dominant share of the angular momentum throughout the relevant stage of the system's evolution:
\begin{align}
\label{AMratio}
\frac{I_{\star} M R_{\star}^2 \omega_{\star}}{m \, \sqrt{\G M a}} \gg 1,
\end{align}
where $I_{\star}$ is the dimensionless moment of inertia, $M$ is stellar mass, $R_{\star}$ is stellar radius, $\omega_{\star}$ is stellar rotation rate, $m$ is the inner planet's mass, and $a$ is the inner planet's semi-major axis. 

The interior structure of a young, fully convective Sun-like star can be approximated by a polytropic body of index $3/2$, which is characterized by $I_{\star} = 0.21$ and a Love number of $k_{2\star} = 0.28$ (e.g. \citealt{BatyginAdams2013}). Adopting a T-Tauri stellar radius of $R_{\star} \sim 2.5 R_{\odot}$, a rotational period of $2 \pi/\omega_{\star} \sim 3$ days, along with typical hot Jupiter parameters of $m \sim 10^{-3} M_{\odot}$ and $a \sim 0.05$ AU, the ratio (\ref{AMratio}) evaluates to $\sim 10$. Our assumed angular momentum hierarchy is thus well satisfied in the regime of interest. Because the planetary angular momentum budget is essentially negligible compared to that of the host star's spin, the stellar spin axis can be taken to be stationary. Without loss of generality, we can orient the coordinate system such that the $\hat{z}-$axis is aligned with the spin pole of the star. 

We further assume that the protoplanetary disk (and the planetary orbits embedded within it) are initially inclined with respect to the stellar spin axis by an angle, $i$. There exist a multitude of processes by which the stellar spin axis may become misaligned with the orbital plane of the disk. One such process is stochastic accretion of disk material during collapse of a molecular cloud core \citep{Bateetal2010,Spaldingetal2014,Fieldingetal2015}. Another mechanism for early excitation of disk-star misalignment is associated with primordial binary stellar companions that exert gravitational torques upon the disk, and thereby excite stellar obliquity \citep{Batygin2012,BatyginAdams2013,Lai2014}. Within the framework of the latter mechanism, it has been shown that the entire possible range of spin-orbit misalignments can be trivially generated \citep{SpaldingBatygin2014}. However, for the purposes of our perturbative treatment, we limit our scope to consideration of small star-disk inclinations, leaving a more general treatment for numerical experiments that will follow. While this assumption is not strictly necessary\footnote{Indeed, we could have in principle chosen to place no restriction on planet-star inclination and expand the Hamiltonian in powers of $(\tilde{a}/a)$. Instead, here we shall place no restrictions on the semi-major axis ratio, but assume small inclinations.}, it does allow for a somewhat simplified treatment of the dynamics. 

The disturbing (non-Keplerian) part of star-planet interactions can be modeled using standard techniques of celestial mechanics. To do this, we first replace the rotationally distorted star with a point mass, orbited by a ring of mass
\begin{align}
\tilde{m} = \left[ \frac{3\, M^2 \omega_{\star}^2 R_{\star}^3 I_{\star}^4}{2 \mathcal{G} k_{2\star} } \right]^{1/3},
\end{align}
and semi-major axis
\begin{align}
\tilde{a} = \left[\frac{4\, \omega_{\star}^2 k_{2\star}^2 R_{\star}^6}{9 I_{\star}^2 \mathcal{G} M} \right]^{1/3}.
\end{align}
These expressions are derived by matching the external gravitational potential and the moment of inertia of the point-ring system to that of a polytropic spheroid. 

Provided that the inner planet does not back-react onto the stellar spin axis, we arrive at the Hamiltonian, $\Ham$, that governs the inner planet's inclination dynamics in the restricted approximation. Accordingly, we define \Poincare\ action-angle variables:
\begin{align}
\label{Poincarefull}
&Z = \sqrt{\G M a} \, \big(1 - \cos(i) \big) &z= -\Omega,
\end{align}
where $\Omega$ is the longitude of the ascending node of the inner planet. The relevant orbit-averaged (i.e. secular) expression, expanded to leading order in mutual inclination then reads \citep{MD99}:
\begin{align}
\label{Hhj}
\Ham &= - \frac{\G \, \tilde{m}}{a}\left[ - \frac{1}{2} \, \tilde{\alpha} \, \tilde{b}_{3/2}^{(1)} \, \sin^2 \left(\frac{i}{2}\right) \right] = \frac{n}{4} \frac{\tilde{m}}{M} \tilde{\alpha} \, \tilde{b}_{3/2}^{(1)} \,Z.
\end{align}
In the above expression, $\tilde{\alpha}=\tilde{a}/a$ and $\tilde{b}_{3/2}^{(1)}$ is the Laplace coefficient, the general form of which is written as:
\begin{align}
\label{Lcoeff}
b_{s}^{(j)} = \frac{1}{\pi} \oint \frac{\cos(j \, \psi)}{(1-2\,\alpha \, \cos(\psi)+\alpha^2)^{s}} \, d \psi.
\end{align}
For reasonable choices of parameters, $\tilde{a} \ll a$, and we are justified in replacing the explicit form of the Laplace coefficient with its leading order hypergeometric expansion $\tilde{b}_{3/2}^{(1)} \simeq 3 \tilde{\alpha}$.

Hamiltonian (\ref{Hhj}) is independent of the angle $z$, rendering $Z$ a constant of motion. Physically, this means that the inclination angle, $i$, is preserved as the orbital plane of the planet precesses about the spin-axis of the host star. The rate of nodal regression, $\nu$ (advance of $z$), is trivially obtained from equation (\ref{Hhj}):
\begin{align}
\label{nu}
\nu = \frac{\partial \, \Ham}{\partial Z} = n \left[ \frac{k_{2\star}}{2} \left( \frac{R_{\star}}{a} \right)^{2} \left( \frac{R_{\star}^3 \, \omega_{\star}^2}{\G \, M} \right) \right].
\end{align}
The specification of the dynamical behavior of the inner planet is now complete. Note that the rate of nodal regression, $\nu$, only depends on stellar parameters, and does not change as a result of planetary accretion. 

Next, we consider the dynamical behavior of the test particle residing on the outer orbit. The observed distribution of multi-planet systems discovered by the \textit{Kepler} mission has shown that orbital commensurabilities do not comprise a dominant fraction of the overall sample \citep{Fabryckyetal2014}. Possible reasons for the relative lack of resonances include turbulent fluctuations in protoplanetary disks \citep{Adamsetal2008,ReinPapaloizou2009}, resonant metastability \citep{GoldreichSchlichting2014,DeckBatygin2015}, as well as small disk eccentricities \citep{Batygin2015}. In light of this fact, we assume that the test particle does not reside on a resonant orbit, and concentrate on secular interactions, whereby all harmonics involving mean longitudes in the Fourier expansion of the Hamiltonian are averaged out \citep{Morbidelli2002}. We note that the averaging process corresponds to a near-identity change of variables that yields semi-major axes that are frozen in time.

Neglecting\footnote{Mixed eccentricity-inclination terms only enter the expansion of the disturbing function at fourth order \citep{EllisMurray2000}.} eccentricities as before, the governing Hamiltonian for the outer orbit is written as follows:
\begin{align}
\label{Htp}
\Hamp &= \frac{\G \, \tilde{m}}{a'}\left[ \frac{1}{2} \, \tilde{\alpha}' \, \tilde{b}_{3/2}^{(1)'} \, \sin^2 \left(\frac{i'}{2}\right) \right] \nonumber \\
&- \frac{\G m}{a'} \bigg[f_1 \, \sin^2 \left(\frac{i'}{2}\right) + f_2 \, s \, \sin \left(\frac{i'}{2}\right) \nonumber \\
&\times \cos \left( \Omega' - \Omega \right) + f_3\, \sin^4 \left(\frac{i'}{2}\right) \bigg],
\end{align}
where
\begin{align}
\label{fsec}
f_1 &= - \frac{f_2}{2} = - \frac{1}{2} \alpha b_{3/2}^{(1)}, \nonumber \\
f_3 &= \frac{3}{16} b_{5/2}^{(-2)} + \frac{3}{4} b_{5/2}^{(0)}+ \frac{3}{16} b_{5/2}^{(2)},
\end{align}
where $\alpha = a/a'$ and for ease of notation, we write $s = \sin (i/2) \ll1$, which we treat as a small parameter.

The first line of equation (\ref{Htp}) mirrors Hamiltonian (\ref{Hhj}) and governs the nodal regression of the test particle orbit, induced by the stellar rotational bulge. The second and third lines represent the averaged gravitational potential arising from the inner planetary orbit. In this expansion, we have retained a term that is fourth-order in $i'$. This term is of substantial importance to the analysis that follows, as it allows for the correct qualitative representation of the topology of the phase-space portrait (see e.g. \citealt{HenrardLamaitre1983}) in the secular resonant regime\footnote{To this end, we could have in principle also retained an additional harmonic term that arises at fourth order in inclination. While this would not destroy the ultimate integrability of $\Hamp$, the additional harmonic does not play a crucial qualitative role in determining the dynamical evolution. Therefore, we have chosen to omit this term to maintain the simplicity of the model.}. 

To proceed further, we must switch to canonically conjugated variables. In direct similarity to equation (\ref{Poincarefull}), we again utilize \Poincare\ variables. However, for the test particle orbit, we scale the action by the specific angular momentum $\sqrt{\G M a'}$:
\begin{align}
\label{Poincarescale}
&Z' = \big(1 - \cos(i') \big) &z'= -\Omega'.
\end{align}
To maintain symplecticity, we must also scale the Hamiltonian by the same factor \citep{LichtenbergLieberman1983}. 

\begin{figure*}[t]
\centering
\includegraphics[width=\textwidth]{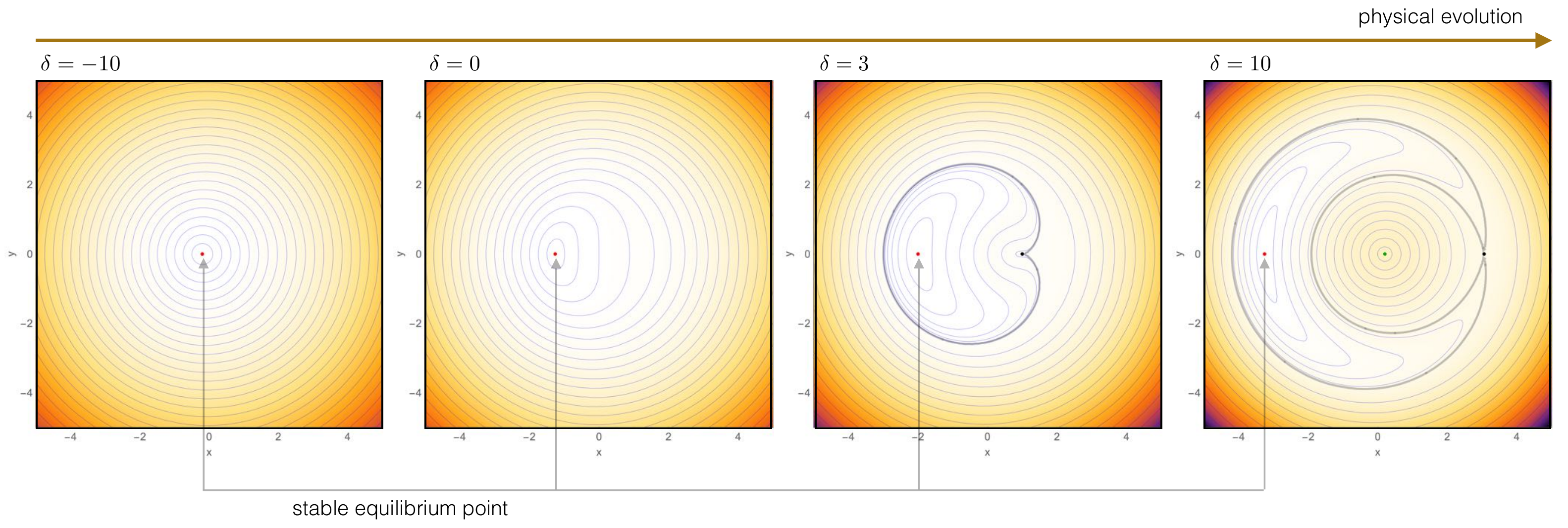}
\caption{Topology of a first-order Andoyer Hamiltonian (equations \ref{Htp4} and \ref{Htp4ecc}) for different values of the proximity parameter, $\delta$. The phase-space portraits are shown in canonical cartesian coordinates $x= \sqrt{2\mathcal{A}}\,\cos(a)$ $y= \sqrt{2\mathcal{A}}\,\sin(a)$ where $\mathcal{A}$ is action and $a$ is angle. The physical evolution of the system (i.e. planetary mass accretion, stellar contraction and spin-down, etc) slowly alters the proximity parameter from $\delta \sim - \infty$ to $\delta \gtrsim 0$. Concurrently, the stable equilibrium point adiabatically shifts away from the origin, implying a steady growth of the action (related to the mutual inclination or the eccentricity). For $\delta \geqslant 3$, a homoclinic curve (along with two additional fixed points) develops in phase-space, allowing for a formal definition of a secular resonance.}
\label{sfmrfig}
\end{figure*}

Recalling from equation (\ref{nu}) that $\Omega = -\nu \, t$, we arrive at the following expression for the Hamiltonian:
\begin{align}
\label{Htp2}
\Hamp &= n' \bigg[ \bigg( \frac{k_{2\star}}{2} \left( \frac{R_{\star}}{a'} \right)^{2} \left( \frac{R_{\star}^3 \, \omega_{\star}^2}{\G \, M} \right) - \frac{f_1}{2} \frac{m}{M} \bigg) Z' \nonumber \\
&-s\, \frac{f_2}{2} \frac{m}{M} \sqrt{2 Z'} \cos \left( z' - \nu \, t \right) \nonumber \\
&- \frac{f_3}{4} \frac{m}{M} \left(Z' \right)^2 \bigg] + \mathcal{T}.
\end{align}
In writing out the above expression, we have made use of the same simplification, as that employed in equation (\ref{nu}). Additionally, we have introduced a dummy action $\mathcal{T}$ conjugate to time, such that $\Hamp$ formally constitutes an autonomous two degree of freedom system \citep{Morbidelli2002}.

Hamiltonian (\ref{Htp2}) is made integrable by a trivial change of variables. Specifically, consider a canonical transformation arising from a type-2 generating function 
\begin{align}
\label{F2}
\mathcal{F}_2 = \left( z' - \nu \, t \right) \Phi + (t) \, \Xi.
\end{align}
The transformation equations yield:
\begin{align}
\label{transfomr}
&Z' = \frac{\partial \mathcal{F}_2}{\partial z'} = \Phi &\phi = z' - \nu \, t, \nonumber \\
&\mathcal{T} = \frac{\partial \mathcal{F}_2}{\partial t} = \Xi - \nu \,\Phi &\xi = t.
\end{align}
By direct substitution, it is clear that $\Hamp$ is now independent of $\xi$, meaning that $\Xi$ is a constant of motion and can therefore be dropped.

Equation (\ref{Htp2}) now takes the form of a first-order Andoyer Hamiltonian, also commonly referred to as the second fundamental model for resonance \citep{HenrardLamaitre1983,Ferraz-Mello2007}:
\begin{align}
\label{Htp3}
\Hamp &= n' \bigg[ \bigg( \frac{k_{2\star}}{2} \left( \frac{R_{\star}}{a'} \right)^{2} \left( \frac{R_{\star}^3 \, \omega_{\star}^2}{\G \, M} \right) - \frac{f_1}{2} \frac{m}{M} - \frac{\nu}{n'}\bigg) \Phi \nonumber \\
&-s\, \frac{f_2}{2} \frac{m}{M} \sqrt{2 \Phi} \cos \left( \phi \right) - \frac{f_3}{4} \frac{m}{M} \Phi^2 \bigg].
\end{align}
Recalling that $R_{\star}$, $\omega_{\star}$, $k_{2\star}$, and $m$ all evolve slowly in time, we are faced with the classical problem of adiabatic dynamical evolution of an integrable Hamiltonian \citep{Kruskal1962,Neishtadt1975,Henrard1993}. 

The three constants that appear in the Hamiltonian are not truly independent of each other. Consequently, to obtain a better sense of the evolutionary regime, we can scale the variables in a way that allows us to introduce a single resonance proximity parameter, $\delta$. Following \citet{Henrard1982,BorderiesGoldreich1984}, we scale the action $\Phi$ and the Hamiltonian itself by a constant factor
\begin{align}
\label{eta}
\eta &= \left(s \, \frac{f_2}{f_3} \right)^{2/3}.
\end{align}
Note that this reduction is only meaningful if $s \neq 0$. 

To complete the preparation of the Hamiltonian, we further scale it by $n'\,f_3\, \eta\, (m/M) /4$, and change the canonical unit of time accordingly \citep{Peale1986}. Equation (\ref{Htp3}) thus becomes
\begin{align}
\label{Htp4}
\Hamp &= \delta\, \Phi - \Phi^2 - 2 \sqrt{2 \Phi} \cos(\phi),
\end{align}
where the expression for $\delta$ reads:
\begin{align}
\label{delta}
\delta &= \left(\frac{64}{f_2^2 \, f_3 \, s^2} \right)^{1/3} \nonumber \\
&\times \left[ \frac{M}{m} \left( \frac{k_{2\star}}{2} \left(\frac{R_{\star}}{a'}\right)^2 \frac{R_{\star}^3 \omega_{\star}^2}{\G M} - \frac{\nu}{n'} \right) - \frac{f_1}{2} \right].
\end{align}
Figure (\ref{sfmrfig}) shows a series of phase-space portraits of $\Hamp$ for different values of $\delta$. 

All of the physical evolution of the system is now encapsulated into the variation of a single parameter, whose limiting values warrant examination. As a first case of interest, consider a regime where $\omega_{\star} > 0$ (also meaning that $\nu>0$) and $m = 0$. Because $a' > a$, substitution of expression (\ref{nu}) for $\nu$ implies that the quantity inside the square brackets in equation (\ref{delta}) is negative. Therefore, for such a parameter choice, $\delta = - \infty$ for any choice of $a'$ and the dynamics are trivial. 

Next, consider a case where $\omega_{\star} > 0$ and $m > 0$. Recalling that $f_1$ is negative definite, and decreases in magnitude with $a'$ (i.e. diminishing $\alpha$), equation (\ref{delta}) dictates that as $a'$ is increased, $\delta$ will range from values above $0$ to values below $0$. As a final example, notice that given sufficient time, stellar rotation will diminish to a degree where the rotational flattening becomes negligible \citep{Bouvier2013}. In this case, we may take $\omega_{\star} = \nu = 0$, which implies that $\delta >0$ for all values of  $a'$. 

As can be inferred from the above discussion, over time the system transitions from a state where $\delta = - \infty$ to 
\begin{align}
\label{deltafin}
\delta \underset{t\rightarrow \infty}{=} -\frac{2 f_1}{\big(f_2^2 \, f_3 \, s^2\big)^{1/3}}>0.
\end{align}
Within the framework of our model, this transition occurs in two stages that come about on different timescales: the first is associated with the growth of $m$, and the second arises from the fading of the stellar quadrupole moment. More generally, it would be naive to expect that the evolution of $\delta$ is linear. In fact, within the context of a more comprehensive model, it need not even be monotonic. Fortunately, however, as long as the parametric evolution of the system occurs on a timescale that is much longer than the dynamical timescale (i.e. adiabatic limit), the initial and final states of the system are dictated only by the endpoints of $\delta$'s evolutionary track, and the path followed in between is not important, as long as the trajectory in question does not encounter homoclinic curves \citep{Cary1986,Henrard1993}.

Formally, the threshold rate of change of $\delta$, below which the adiabatic condition is satisfied, can be defined as follows: the timescale on which $\dot{\delta}$ carries the unperturbed trajectory across the width of a resonance (defined as the difference between the maximal and minimal excursions of $\Phi$ on a separatrix) must be longer than the resonant libration period \citep{Friedland2001}. This condition is satisfied for reasonable choices of parameters for the system at hand, other than for pathologically small values of $s$ (e.g. $s = 10^{-4}$).

Taking advantage of the system's adiabatic nature, we may define the following quasi-integral of motion \citep{Neishtadt1975}:
\begin{align}
\label{J}
\mathcal{J} = \oint \Phi \, d \phi,
\end{align}
where the integral is taken along the trajectory's path in phase-space. Physically, the near-conservation of $\mathcal{J}$ implies that the phase-space area engulfed by the orbit is preserved in face of the variation of $\delta$.

At the inception of the system's dynamical evolution, when $\delta = - \infty$, all trajectories encircle an elliptic equilibrium of the system that resides at $\Phi = 0$. As $\delta$ grows to values above $\delta \gtrsim 0$, this fixed point is advected\footnote{Note that at a new equilibrium of the Hamiltonian (\ref{Htp4}) appears at $\delta = 3$ and bifurcates into a hyperbolic equilibrium and an elliptic equilibrium above this critical value.} to larger values of $\Phi$.  Correspondingly, any small patch of phase space area that engulfed this equilibrium is also carried to high inclinations, as the system becomes locked in a secular resonance. This result is key to understanding how orbit-orbit misalignments are excited in close-in planetary systems. 

Suppose that the initial orbit of the test particle lies in the same plane as that of the inner orbit. By virtue of the assumption of the smallness of $s$ and the adiabatic principle discussed above, we immediately arrive at the (strictly real) equilibrium of the Hamiltonian as a good approximation for the end-state action attained by the outer orbit. One way to compute this action is to derive an equation for the equilibrium value of $\Phi$ from equation (\ref{Htp4}) as a function of $\delta$, for which we can plug in expression (\ref{deltafin}) and subsequently rescale the answer by the constant factor $\eta$ (given by equation \ref{eta}). An equivalent (and arguably simpler) approach would be to derive the equilibrium condition directly from Hamiltonian (\ref{Htp3}), setting $\omega_{\star} = \nu = 0$ and\footnote{The signs upfront the kinetic terms in Hamiltonian (\ref{Htp4}) dictate that the resonant equilibrium point lies at $\phi=\pi$. Flipping these signs would change the location of the equilibrium to $\phi=0$.} $\phi = \pi$. Accordingly, we have:
\begin{align}
\label{equilib}
\frac{\partial \Hamp}{\partial \Phi} = \frac{n'}{2} \frac{m}{M} \left(s \frac{f_2}{\sqrt{2 \Phi}} - f_1 - f_3 \Phi \right) = 0.
\end{align}

Equation (\ref{equilib}) admits a closed-form solution for $\Phi$. However, the solution is rather cumbersome, so it makes sense to once again utilize the assumed smallness of $s$ to expand the expression as a Taylor series and obtain a first-order approximation for the final inclination of the outer orbit:
\begin{align}
\label{equilibi}
i' &= \arccos\left(\frac{f_1 + f_3}{f_3} \right) - s \, \frac{f_2}{f_1} \nonumber \\
&\times \left(\sqrt{4+ 2\frac{f_1}{f_3}} \right)^{-1} +\, \mathcal{O}(s^2).
\end{align}
To leading order, this solution does not depend on $s$. This may appear counter-intuitive, since a null $s$ should produce a null $i'$. To resolve this discrepancy, recall that equation (\ref{equilib}) is derived in the adiabatic approximation, and setting $s=0$ removes the harmonic term from the Hamiltonian. In other words, $s=0$ corresponds to an infinite resonant libration period, which violates the adiabatic condition by construction. Consequently, the solution (\ref{equilibi}) strictly applies in the regime where $0< s \ll 1$. 

There is another striking feature of the solution (\ref{equilibi}), and that is the actual final value of $i'$. The functional forms of the coefficients presented in expression (\ref{fsec}) are in general quite complicated and must be evaluated numerically. However, if we restrict ourselves to orbits characterized by $\alpha$ much smaller than unity, we may expand these coefficients as hypergeometric series (as already done before). Upon doing so, we obtain $f_3 = - f_1 = 3 \alpha^2/2$. This means that to leading order in $s$, 
\begin{align}
\label{iprimefin}
i' \rightarrow \arccos(0) = \pi/2.
\end{align}
The theoretical analysis presented herein suggests that \textit{the process of resonant excitation of orbital inclination ultimately leads to orthogonal orbits}. Below, we examine how this assertion fares when confronted with direct numerical tests.

To conclude this section, we summarize the qualitative content of our solution. By utilizing classical perturbation theory, we have shown that the physical processes of mass accretion by an interior secondary orbit and the disappearance of the primary body's quadrupolar gravitational potential, can excite substantial orbit-orbit misalignments. This comes about as a consequence of adiabatic evolution into a first-order secular inclination resonance. That is, the system starts out in a state where the nodal regression of the inner orbit is faster than that of the outer orbit, simply because it is closer to the host star. As the inner body gains mass and the star spins down, the nodal regression of the outer orbit accelerates, and eventually a secular commensurability is encountered. As a way to maintain the commensurability during subsequent evolution, the outer orbit's inclination grows, which in turn reduces the forced rate of nodal regression due to the presence of the non-linear action term in the Hamiltonian. Correspondingly, by the time the quadruplolar component of the stellar potential completely disappears, the nodal regression of the inner orbit ceases and the outer orbit finds itself in an orthogonal state, such that the orbit-projected torque vanishes and its nodal regression also stalls.

There exist a number of caveats that come into light upon closer examination. First, the described process only works if the more massive body resides on the inner orbit. Indeed, accretion of mass by the outer planet (while keeping the inner body a test-particle) will result in a configuration where the nodal recession of the inner orbit is always faster, meaning that a secular resonance can never be established. Second, we have neglected the presence of the disk. Gravitational interactions of the system with a massive disk will lead to an enhanced nodal regression of the bodies during the disk-bearing phase of the stellar lifetime \citep{Ward1981}, as well as damping of mutual inclination \citep{PapaloizouLarwood2000}. While both of these complications affect our calculations on a detailed level, we show below that the envisioned picture generally holds.

We note that orbital excitation by sweeping secular resonances has previously been studied within the context of the solar system (see e.g. \citealt{Ward1976,Ward1981,NagasawaIda2000,NagasawaIda2002,MintonMalhotra2011,AgnorLin2012}). In contrast with our work, these studies do not observe capture, and instead report impulsive excitation of the orbital parameters upon encounter. However, these analyses are consistently based on a second order (Lagrange-Laplace) expansion of the Hamiltonian and therefore cannot properly model the establishment of secular commensurabilities, because homoclinic curves do not exist in phase space within this framework, disallowing a resonant domain to even be defined. A counter-example is provided by the $N$-body simulations of \citet{Nagasawa2005}, where long-term capture into first order secular eccentricity resonances is observed.

A final point worthy of attention is that as the host star spins down, the assumed angular momentum hierarchy (equation \ref{AMratio}) breaks down. Therefore, the assumption that the stellar spin-axis remains fixed in inertial space does not formally apply towards the end the evolutionary track of $\delta$. If $m$ is taken to approach a typical hot Jupiter-like mass, however, it can be shown that by the time the angular momentum hierarchy is severely violated, the value of $\delta$ is already very close to that given by equation (\ref{deltafin}). Accordingly, this complication does not pose a practical limitation.

\subsubsection{Eccentricity Evolution}

\begin{figure}[t]
\centering
\includegraphics[width=\columnwidth]{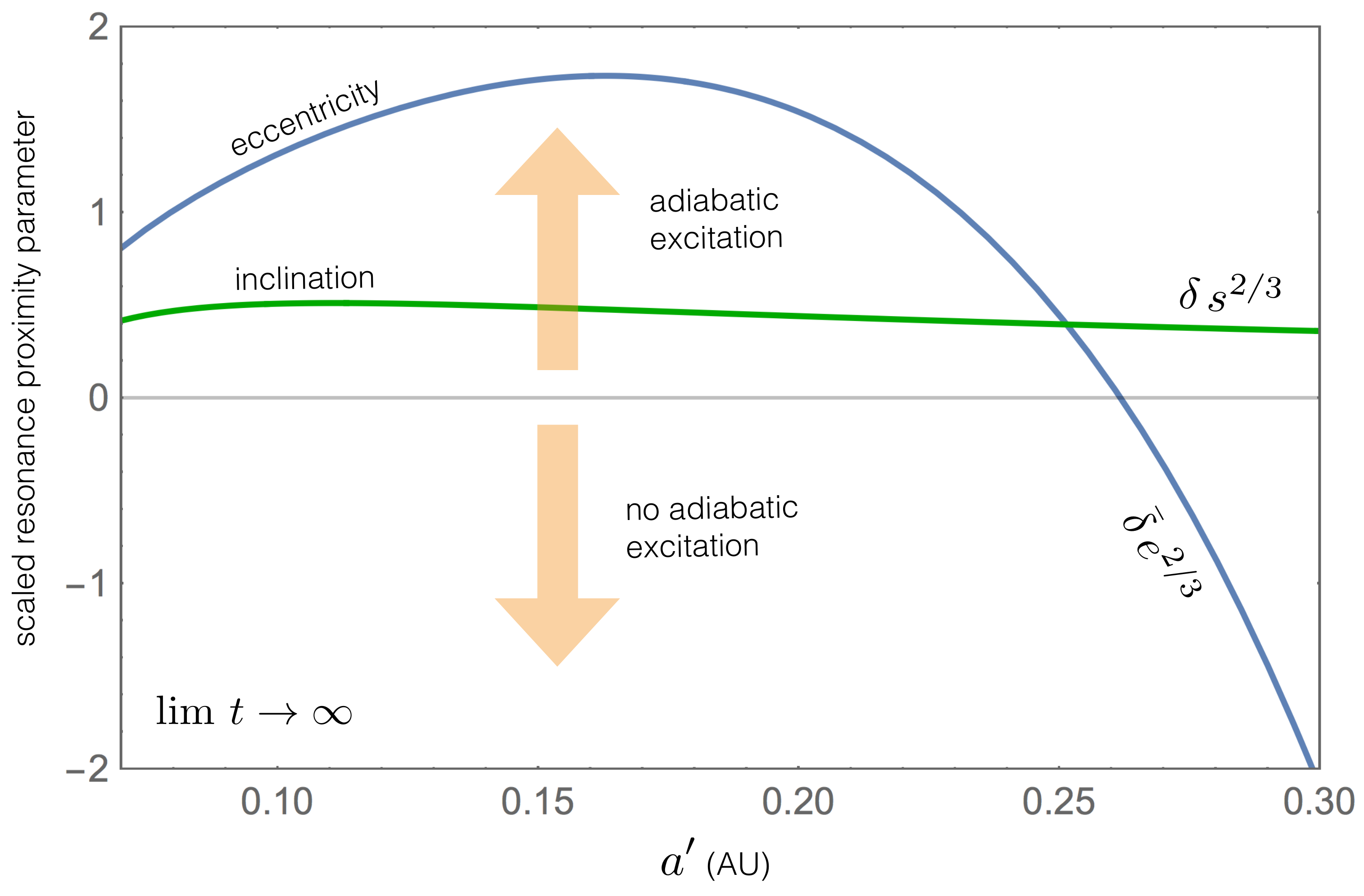}
\caption{The scaled resonance proximity parameters $\delta \, s^{2/3}$ and $\bar{\delta} \, e^{2/3}$ in the limit of $t\rightarrow \infty$ (i.e. a regime where the relevant physical evolution of the system has concluded) as functions of test particle semi-major axis, $a'$. Note that because of the smallness of $s$ and $e$, the real proximity parameters are strongly amplified in comparison to the values shown on the ordinate in this Figure. This implies that any value above $0$ signals large excitation, and any value below $0$ corresponds to null adiabatic growth. Owing to relativistic and tidal precession of the hot Jupiter's apsidal line, the secular eccentricity resonance is detuned beyond $a'\gtrsim 0.27$ AU. There is no corresponding effect that affects the inclination degree of freedom. However, it is important to simultaneously keep in mind that beyond a critical $a'$, the adiabatic limit (which is set by the physical evolution timescales) will be broken and no excitation of the action will take place, even if the proximity parameter evolves to positive values.}
\label{deltafig}
\end{figure}

Thus far, we have neglected orbital eccentricities in order to obtain an integrable Hamiltonian. Recalling that eccentricity and inclination terms are decoupled to second order in either quantity \citep{MD99}, this assumption is justified as long as the eccentricities remain small. However, the ultimate result of adiabatic capture into a secular inclination resonance raises the question of whether or not similar dynamics can occur in the degree of freedom associated with the outer orbit's eccentricity. Fortunately, very similar techniques to those already employed above can be used to attack this problem.

In direct analogy to equation (\ref{nu}), the inner orbit's longitude of perihelion will not remain stationary in time, and will instead precess at a constant rate: $\varpi = \mu\, t$. In contrast with nodal regression, however, the rate of apsidal precession comprises numerous contributions \citep{RagozzineWolf2009}:
\begin{align}
\label{mu}
\mu &= n \bigg[ \frac{k_{2\star}}{2} \left( \frac{R_{\star}}{a} \right)^{2} \bigg( \frac{R_{\star}^3 \, \omega_{\star}^2}{\G \, M} \bigg) + \frac{k_{2p}}{2} \left( \frac{R_{p}}{a} \right)^{5} + 3 \, \frac{\G \, M}{a \,c^2} \nonumber \\
&+ 15 \,\bigg( \frac{k_{2\star}}{2} \frac{m}{M} \left( \frac{R_{\star}}{a} \right)^{5}+ \frac{k_{2p}}{2} \frac{M}{m} \left( \frac{R_{p}}{a} \right)^{5} \bigg)  \bigg],
\end{align}
where quantities with the subscript $p$ refer to the planet, and $c$ is the speed of light. For simplicity, in the above expressions we have assumed that the planetary obliquity is null (and therefore only contributes to apsidal precession and not nodal regression), the rotation rate is synchronous with the orbital frequency \citep{Hut1981}, and that eccentricities are sufficiently low to warrant the retention of only leading-order terms in $e$ (see \citealt{Kopal1978} for generalized equations). The physical origin of the first two terms on the first line of equation (\ref{mu}) is rotational distortion; the last term on the first line governs apsidal advance facilitated by general relativity; terms on the second line stem from tidal deformation of the bodies. 

Quantitatively, there is a hierarchy among the various terms in equation (\ref{mu}) that is strongly dependent on the evolutionary state of the system. For typical hot Jupiter parameters of $R_{p} \sim 1.3 R_{\rm{Jup}}$, and $k_{2p} \sim 0.3$ (see e.g. \citealt{Batyginetal2009,Mardling2010}), the tidal and relativistic terms are comparable (to an order of magnitude) and greatly exceed the planetary rotational contribution. However, because the rotational period of T-Tauri stars is approximately commensurate with orbital periods of hot Jupiters, the contribution due to the stellar rotational bulge dominates other terms during the pre-main-sequence stage of stellar lifetime. This is in stark contrast with later stages of main-sequence evolution, where relativity and the planetary tidal contributions dominate over all other terms \citep{RagozzineWolf2009}. 

In order to obtain a qualitative description of the eccentricity dynamics, it is useful to first ignore inclination evolution and consider a Hamiltonian that governs purely planar interactions (the fully coupled system will be treated numerically later). Disregarding the test particle's physical structure, we have \citep{MD99}:
\begin{align}
\label{Hecc}
\Hampb &= - \frac{\G \, \tilde{m}}{a'}\left[ \frac{1}{8} \, \tilde{\alpha}' \, \tilde{b}_{3/2}^{(1)'}  (e')^2 \right] - \frac{3}{2} \left( \frac{\G M}{a' c} \right)^2 (e')^2 \nonumber \\  
&- \frac{\G m}{a'} \bigg[ \bar{f}_1 (e')^2  + \bar{f}_2 \, e \, e' \cos(\varpi' - \varpi) + \bar{f}_3 (e')^4 \bigg], 
\end{align}
where the over-bar is used to distinguish corresponding expressions for eccentricity and inclination and $e \ll 1$ is the inner planet's eccentricity. The newly introduced interaction coefficients read:
\begin{align}
\label{fbar}
\bar{f}_1 &= \frac{1}{8} \alpha b_{3/2}^{(1)} \nonumber \\
\bar{f}_2 &= -\frac{1}{4} \alpha b_{3/2}^{(2)} \nonumber \\
\bar{f}_3 &=  \bigg[\frac{3}{16} \alpha \mathcal{D}_{\alpha} + \frac{9}{32} \alpha^2 \mathcal{D}_{\alpha}^2 + \frac{3}{32} \alpha^3 \mathcal{D}_{\alpha}^3 \nonumber \\
&+ \frac{1}{128} \alpha^4 \mathcal{D}_{\alpha}^4 \bigg] b_{1/2}^{(0)}
\end{align}
where $\mathcal{D}_{\alpha} = \partial/\partial \alpha$.


Introducing scaled canonical coordinates akin to equations (\ref{Poincarescale}) and (\ref{transfomr})
\begin{align}
\Psi &= \big(1 - \sqrt{1-(e')^2} \big)\simeq \frac{(e')^2}{2} \nonumber \\
\psi &= \mu \, t - \varpi',
\label{Psi}
\end{align}
we may rewrite the Hamiltonian in the following way:
\begin{align}
\label{Hecc2}
\Hampb &= n' \bigg[\bigg( \frac{\mu}{n'} - \frac{k_{2\star}}{2} \bigg( \frac{R_{\star}}{a'} \bigg)^2 \bigg( \frac{R_{\star}^3 \, \omega_{\star}^2}{\G \, M} \bigg) - 3  \frac{\G M}{a' c^2} \nonumber \\
 &- 2 \bar{f}_1 \frac{m}{M} \bigg) \Psi - 4 \bar{f}_3 \frac{m}{M} \Psi^2 - \bar{f}_2 \, e \, \sqrt{2 \Psi} \cos(\psi) \bigg].
\end{align}
Because the functional form of this expression is in essence identical to that of Hamiltonian (\ref{Htp3}) and the parameters evolve in the same sense, the transformation of the phase-space portrait of this Hamiltonian will follow the same sequence as that depicted in Figure (\ref{sfmrfig}). In particular, scaling the actions by the factor\footnote{Note that similarly to the scaling (\ref{eta}), this procedure is only sensible if the inner planet's eccentricity is finite.}
\begin{align}
\label{etabar}
\bar{\eta} &= \frac{1}{4} \left(e \, \frac{\bar{f}_2}{\bar{f}_3} \right)^{2/3},
\end{align}
and introducing a reduced dimensionless time $4\, t\, n'\,\bar{f}_3\, \bar{\eta}\, (m/M)$, we obtain the same functional form for the Hamiltonian as equation (\ref{Htp4}):
\begin{align}
\label{Htp4ecc}
\Hamp &= \bar{\delta}\, \Psi - \Psi^2 - 2 \sqrt{2 \Psi} \cos(\psi).
\end{align}

As before, we have the following expression for the proximity parameter in the limit where planetary accretion and stellar spin-down are complete:
\begin{align}
\label{deltabarfin}
\bar{\delta} &\underset{t\rightarrow \infty}{=} \frac{1}{\big(\bar{f}_2^2 \, \bar{f}_3 \, e^2\big)^{1/3}} \bigg[2 \bar{f}_1  + 3 \frac{M}{m} \frac{\mathcal{G} \, M}{a \, c^2} - \frac{\mu}{n'} \frac{M}{m} \bigg].
\end{align}
When evaluating the above equation, we must also set $\omega = 0$ in equation (\ref{mu}). Figure (\ref{deltafig}) depicts $\delta$ and $\bar{\delta}$ as functions of $a'$ in the limit of $t\rightarrow \infty$. 

An interesting feature that highlights the difference between the degrees of freedom related to the eccentricity and inclination is that unlike the final result for $\delta$ (given by equation \ref{deltafin}) which is positive for all $a'$, $\bar{\delta}$ transitions from positive to negative beyond a critical $a'\simeq 0.27$ AU. Qualitatively, this occurs because given a sufficient orbital separation, relativistic and tidal contributions to the apsidal precision render it impossible for a distant test particle to precess at a comparable rate to that of the hot Jupiter, effectively detuning the resonance \citep{AdamsLaughlin2006}.

Generally, the presented argument suggests that \textit{capture into a secular eccentricity resonance can indeed occur, provided that the inner orbit is not circular}. We can in principle proceed to calculate the extent to which resonant excitation will adiabatically enhance the outer object's eccentricity. It bears noting, however, that such an estimate may be meaningless, given that as eccentricities grow adiabatically, close encounters will give way to a large-scale instability, thereby transforming the system. 

Taken together, our results suggest that early evolution of close-in planetary systems concludes either in gross orbit-orbit misalignments, or large-scale instability, depending on whether or not the hot Jupiter orbit remains circular. Accordingly, we now turn to numerical simulations to confirm and extend our ideas.

\subsection{N-Body Simulations}
We adopt a standard $N$-body gravitational dynamics solver, augmented to account for relativistic effects, rotational and spin-axis evolution, as well as the associated quadrupolar fields of the bodies. The specific corrections to pure Newtonian gravity were implemented using the formulae presented below.

\begin{figure}
\centering
\includegraphics[width=\columnwidth]{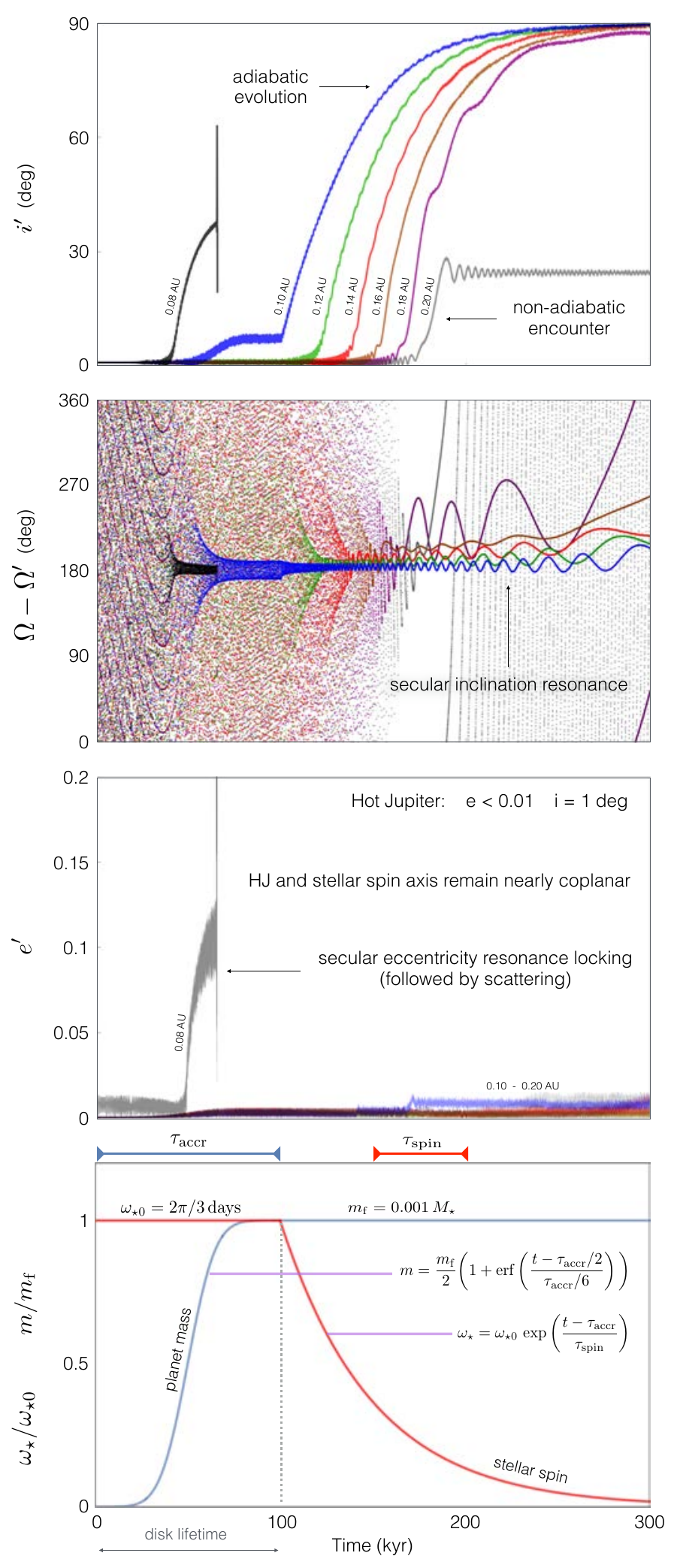}
\caption{Dynamical evolution of the test particle obtained within the nominal suite of $N$-body simulations. Here the hot Jupiter orbit is taken to be circular and $i=1\deg$. The first, second, and third panels from the top respectively show inclination, resonance angle, and eccentricity evolution for test particles with $a'=0.08 - 0.2$ AU. Analytically described adiabatic growth of inclination (with the exception of $a'=0.2$ AU case where the adiabatic limit is artificially broken) and eccentricity (in the $a'=0.08$ AU case) is well reproduced by direct numerical simulations. Throughout the calculations, the orbit of the hot Jupiter and the stellar spin-axis remain nearly coplanar at $i\simeq 0$. A graphical representation of parameterizations of planetary growth and stellar spin-down, employed in the $N$-body simulations is shown in the bottom panel.}
\label{inc1fig}
\end{figure}

General relativistic advance of the planetary periapse is mimicked by introducing accelerations arising from an additional term of the stellar potential of the form
\begin{align}
\label{GR}
\mathbf{a}_{\rm{GR}} = \nabla \bigg[ 3\left( \frac{\G   M}{c\, r} \right)^2 \bigg],
\end{align}
where $r$ is the star-planet separation. This correction yields the exact secular relativistic precession rate, at the expense of a small $(\mathcal{O}(m/M)^2)$ error in the mean motions \citep{NobiliRoxburgh1986}. 

The acceleration arising from the quadrupolar component of the gravitational field of body $i$ experienced by body $j$ takes the form \citep{MardlingLin2002}:
\begin{align}
\label{aQ}
\mathbf{a}_{\rm{Q},j} &= \frac{k_{2,i}}{2} \frac{R_{i}^5}{r^4}\bigg[\bigg( 5 (\bm{\omega}_i \cdot \hat{\bm{r}})^2 - \bm{\omega}_i \cdot  \bm{\omega}_i - 12 \frac{G \, m_j}{r^3} \bigg) \hat{\bm{r}} \nonumber \\
&- 2 (\bm{\omega}_i \cdot \hat{\bm{r}})\, \bm{\omega}_i \bigg] \left(1+\frac{m_j}{m_i} \right).
\end{align}
Only quadrupolar planet-star interactions were implemented to save computational costs, as quadrupolar planet-planet interactions are completely negligible in the regime of interest \citep{Correiaetal2015,BatyginMorbidelli2015}. For the inner planet, a typical hot Jupiter radius of $R = 1.3 R_{\rm{Jup}}$ was adopted, along with a Love number of $k_2 = 0.3$. Additionally, we adopted the same stellar parameters as those quoted above. 

The spin-axis evolution of the bodies was computed from:
\begin{align}
\label{spinupdate}
I_j \dot{\bm{\omega}}_{j} &= \frac{m_i \, m_j}{m_i + m_j} \left(\bm{r} \times \mathbf{a}_{\rm{Q},j} \right).
\end{align}
All calculations were performed using the Bulirsch-Stoer algorithm \citep{Pressetal1992}. Tidal evolution was neglected for simplicity, and the planetary spins were initialized to their respective orbit-synchronous values. 

For the purposes of our numerical experiments, we chose to model the accretion of mass by the inner planet and stellar spin-down as subsequent processes. While this separation may be justified from physical grounds, we note that as long as the adiabatic condition is well satisfied for both of the characteristic timescales, it would not have been problematic to model them as occurring simultaneously (since qualitatively they both simply facilitate the evolution of the proximity parameter). 

The accretion of mass was parameterized using the following functional form:
\begin{align}
\label{maccretion}
m = \frac{m_{\rm{f}}}{2} \bigg(1+\erf \left( \frac{t-\tau_{\rm{accr}}/2}{\tau_{\rm{accr}}/6} \right) \bigg),
\end{align}
where $m_{\rm{f}}$ is the final planetary mass (here, chosen to be $m_{\rm{f}}=10^{-3} M_{\odot}$), and $\tau_{\rm{accr}}$ is the accretion time. Although the disk is not modeled directly in these simulations, $\tau_{\rm{accr}}$ can also be nominally thought of as the disk lifetime. Stellar spin-down was implemented using simple exponential decay with a characteristic timescale $\tau_{\rm{spin}}$, that only operated at times exceeding the accretion time:
\begin{align}
\label{spindown}
\dot{\bm{\omega}} &= - \frac{\bm{\omega}}{\tau_{\rm{spin}}} & t > \tau_{\rm{accr}}.
\end{align}
Initial stellar rotation frequency was set to $2\pi/\omega_0=3\,$ days. A graphical representation of these parameterizations is presented in the bottom panel of Figure (\ref{inc1fig}).

For our nominal set of calculations, the planets were initialized with identical inclinations relative to the star's spin ($i'=i=1\deg$), on nearly-circular orbits ($e\lesssim 10^{-3}$) with randomized nodal lines. Mean anomalies and longitudes of perihelia were also given random values. With respect to physical evolution, we adopted short, but nevertheless adiabatic timescales of $\tau_{\rm{accr}} = 2\, \tau_{\rm{spin}} = 10^5$ years. The integrations spanned $3\times10^5$ years, approximately an order of magnitude shorter than typical disk lifetimes \citep{Haischetal2001}. 

Our chosen timescales are not intended to be representative of real physical values, and are made artificially short to decrease integration time. In reality, the process of giant planet conglomeration takes upwards of $\sim 1\,$Myr (as depicted in Figure \ref{pollackfig}), while stellar evolution generally proceeds on even longer timescales \citep{Bouvier2013}. However, the adiabatic nature of dynamical evolution ensures that the long-term behavior of the system scales with the adopted rates of physical evolution. Therefore, our results easily translate to longer integrations. 

The calculated evolutionary tracks are presented in Figures (\ref{inc1fig}) for various choices of test particle semi-major axis, spanning $0.08-0.2$ AU. Specifically, the first, second, and third panels from the top depict the orbital inclination of the test particle, the difference in the longitudes of ascending node of hot Jupiter and test particle orbits (i.e. the secular harmonic angle that appears in Hamiltonian \ref{Htp}), and the test particle eccentricities respectively. Upon examination, it is immediately clear that the results of the simulations follow our theoretical expectations. Specifically, for all test particle orbits with $a\geqslant0.1$ AU, the eccentricities remain close to zero throughout the evolutionary sequence, and large orbital inclinations are excited. 

The gradual increase in the orbital inclination tracks the stable equilibrium point emphasized in Figure (\ref{sfmrfig}). Moreover, in agreement with equation (\ref{iprimefin}), once secular resonance is established (as can be deduced from the transition towards libration of the critical angle in the middle panel), the test particle orbits tend to an orthogonal state. Note that beyond $a' \gtrsim 0.1$ AU, inclination excitation arises entirely from stellar spin down, and not from the conglomeration of the inner planet. This means that substantial orbit-orbit misalignments are likely excited after the dissipation of the gaseous nebula.  

\begin{figure}
\centering
\includegraphics[width=\columnwidth]{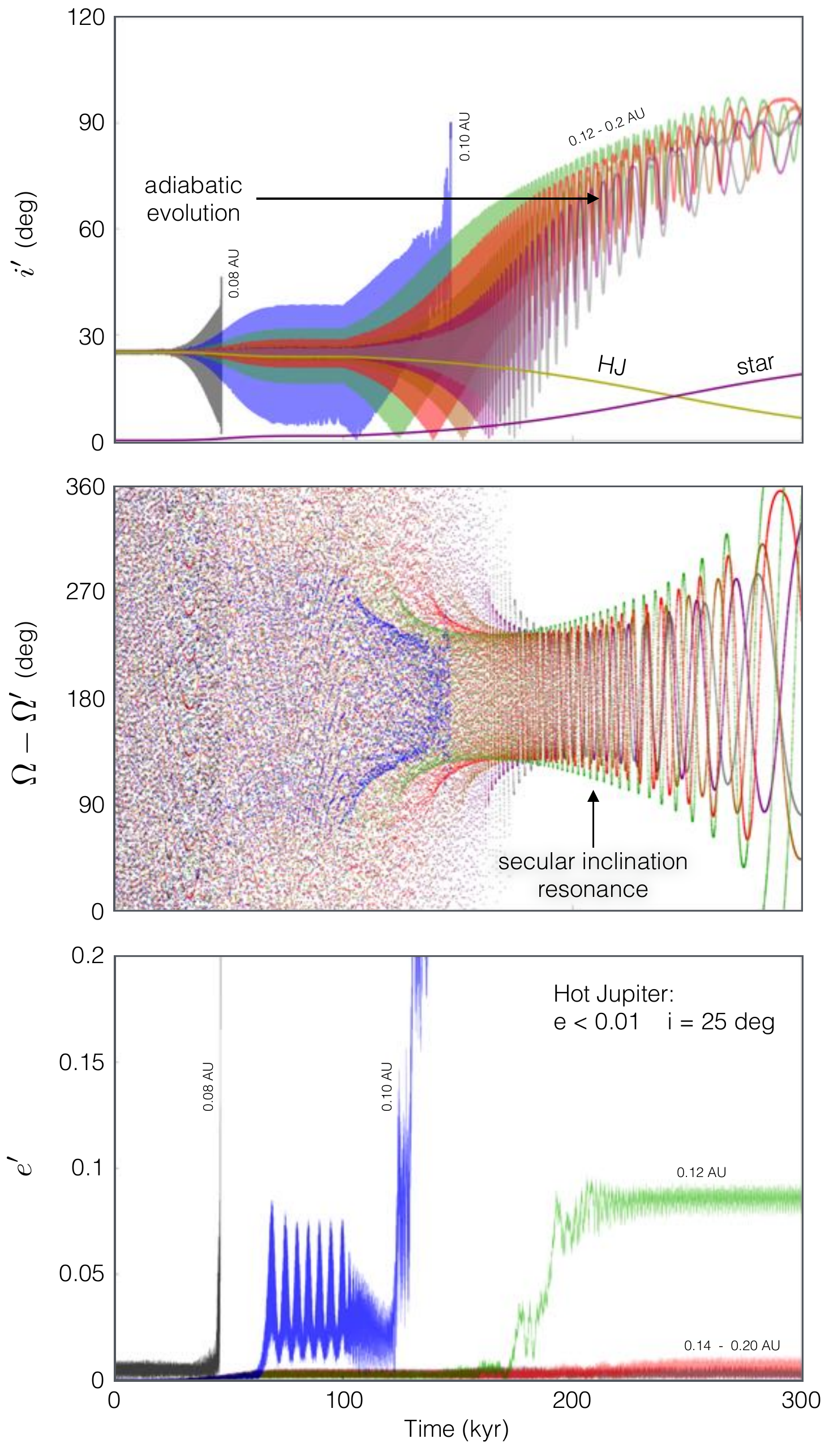}
\caption{Same as Figure (\ref{inc1fig}) but with $i =25\deg$. Note that although the qualitative behavior is similar, the resonant libration amplitude observed in this set of calculations is substantially bigger and eccentricity dynamics exhibited by test particles with $a' \leqslant 0.12$ AU are non-trivial. In the top panel of the Figure, two curves labeled ``HJ" and ``star" depict the orbital inclination of the hot Jupiter and the obliquity of the host star respectively.}
\label{inc25fig}
\end{figure}

\begin{figure*}[t]
\centering
\includegraphics[width=\textwidth]{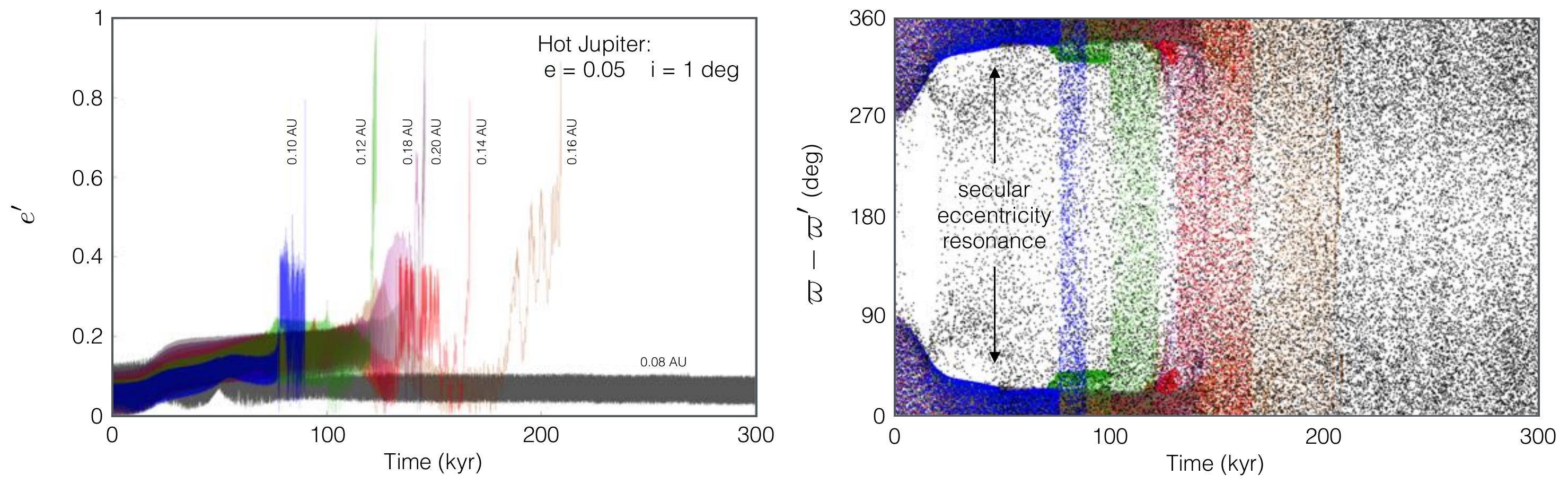}
\caption{Eccentricity evolution of the test particle in a regime where the hot Jupiter orbit is taken to be mildly eccentric (i.e. $e=0.05$). Note that the right panel now depicts the resonant harmonic relevant to the eccentricity degree of freedom (Hamiltonian \ref{Hecc}). In all but one case ($a'\geqslant0.1$AU), large eccentricities are excited by secular resonance locking and the system is transformed by large-scale instabilities.}
\label{eccfig}
\end{figure*}

Notable exceptions within the simulation suite include the experiments where the test particle resides on orbits with $a'=0.08$ AU and $a' = 0.2$ AU. In the former case, a secular eccentricity resonance ensues, resulting in adiabatic eccentricity pumping that in turn leads to scattering. In the latter case, the system remains stable, but the excitation of mutual inclination is stalled due to a violation of the adiabatic limit. To this end, we note that this change in behavior is an artifact of short parametric evolution timescales, and in reality, the adiabatic regime likely extends to somewhat further orbits than $a'=0.2$ AU.

In addition to the nominal series of calculations presented in Figure (\ref{inc1fig}), we also considered numerical experiments where the hot Jupiter was initialized with a significant inclination\footnote{We note that in this case, the randomization of the node will already yield a large misalignment of the the test particle orbit with respect to the hot Jupiter. Nevertheless, it is interesting to examine how the qualitative mode of evolution changes when the assumption of a small $s$ is abandoned.} relative to the stellar spin-axis ($i=25 \deg$) and an enhanced eccentricity ($e=0.05$). In the case of the high initial inclination experiments (Figure \ref{inc25fig}), the evolutionary sequences exhibited qualitatively similar behavior to the nominal array of simulations. In particular, stable adiabatic capture into a secular inclination resonance (resulting in nearly orthogonal orbits) occurred for test particles with $a' \geqslant 0.12$ AU, although the libration amplitude was notably larger (owing to a larger phase-space area initially occupied by the orbits - see equation \ref{J}). Objects with $a' = 0.08$ and $a'=0.10$ AU became violently unstable as a consequence of locking into a secular eccentricity resonance, while similar eccentricity growth for the $a'=0.12$ AU orbit proved short-lived, leaving behind a dynamically excited, but nonetheless stable system. It is further worth noting that in this case, the spin-axis of the star and the inclination of the hot Jupiter itself change adiabatically, as the hierarchy inherent to equation (\ref{AMratio}) begins to break down.

A dramatically different outcome emerged for simulations where the hot Jupiter underwent conglomeration on a mildly eccentric orbit. Particularly, as shown in Figure (\ref{eccfig}), in all but one case ($a'=0.08$ AU) the orbits became violently unstable. Such evolution is in line with the analytic calculation presented in Figure (\ref{deltafig}) which suggests that provided a finite hot Jupiter eccentricity, adiabatic capture into a secular eccentricity resonance is very likely for the orbital range of interest. To this end, note that the angle depicted in Figure (\ref{eccfig}) is that corresponding to the eccentricity resonance (see equation \ref{Hecc}) as opposed to the inclination resonance, as depicted in Figures (\ref{inc1fig}) and (\ref{inc25fig}). Cumulatively, these results point to a distinct possibility that relaxation of the circular constraint on the growing planet's trajectory often leads to the destruction of planetary bodies residing on exterior nearby orbits.

\subsection{Disk Driven Evolution}

The agreement between our analytic and numerical approaches shows that secular theory based on a literal expansion of the disturbing Hamiltonian is well suited for analysis of the problem at hand. We can thus revert to perturbation theory and explore how the model outlined so far is altered within the context of a more complete physical description. Specifically, we want to simultaneously consider the effects of planet-disk interactions and PMS stellar evolution. 

Given that here we implicitly consider protoplanetary disks that are massive enough to spawn a giant planet, it may appear inconsistent to neglect disk-driven migration. It is generally believed that disk-driven migration affects sub-Jovian and Jovian planets in somewhat different regimes (entitled type-I and type-II respectively), where the latter operates on approximately the viscous timescale and the former is thought to be a more rapid process. Both of these processes can affect the evolution of semi-major axis, eccentricity, and inclination. However, for the purposes of our exploratory study, we only consider the effects of direct damping and not radial evolution, for the following reasons.

For type-I migration, state-of-the-art protoplanetary disk models (e.g. \citealt{Bitsch2015} and the references therein) suggest that entropy gradients in the inner nebula can halt or even reverse the migration direction (see also \citealt{2015Natur.520...63B}). Although the details of such models can be exceedingly complex, the existence of a subset of mean-motion resonances among close-in sub-Jovian planets\footnote{Examples of such systems include Kepler-79 and Kepler-223.} points to the fact that type-I migration (however it might proceed) does not ubiquitously drive planets into the magnetospheric cavity, as predicted within the framework of simplified isothermal calculations (e.g. \citealt{Ward1997,Tanaka2002}). In light of this, incorporation of any prescription for radial migration of already close-in planets would introduce unfounded complexity into our model, without providing any additional insight.

For type-II migration, the picture may be somewhat more rudimentary, since gap-opening planets \citep{Cridaetal2006,DuffellMacFadyen2013,Fung2014} simply get carried inwards by the accretionary flow \citep{2007Icar..191..158M}. To this end, we hold no objection to limited radial transport of a giant planet that experienced in-situ formation somewhere in the inner nebula. However, because the rate of radial transport associated with viscous accretion increases with decreasing orbital distance, such an object is likely to rapidly find itself in the inner-most regions of the disk, reproducing initial conditions similar to those considered above. 

With the above arguments in mind, we consider a simplified description of the system where the semi-major axes remain invariant, the hot Jupiter is retained on a circular orbit, and the exterior low-mass planet embedded within the gas is affected by disk torques. For a circular, inclined orbit of the exterior body, disk effects are two-fold. First, the gravitational potential of a rigid, axisymmetric disk induces a nodal regression of an embedded body \citep{Heppenheimer1980,Hahn2003}:
\begin{align}
\label{nodedisk}
\frac{d \Omega'}{dt} \bigg|_{\rm{d}}&= - n' \frac{\pi}{\beta} \bigg( \frac{\Sigma \, (a')^2}{M} \bigg) \simeq \nonumber \\
&-  n' \frac{1}{2\, \beta} \bigg(\frac{m_{\rm{d}}}{M} \bigg) \left( \frac{a'}{a_{\rm{d}}} \right),
\end{align}
where $\beta$ is the aspect ratio of the disk, $\Sigma \propto a^{-1}$ is the gas surface density at semi-major axis $a'$, while $m_{\rm{d}}$ and $a_{\rm{d}}$ are disk mass and size respectively. Second, generation of bending waves by the planetary orbit results in exponential damping\footnote{In addition to damping, bending waves also induce nodal regression. However, the rate tends to be smaller than that given by expression (\ref{nodedisk}) for low-mass planets, and can thus be neglected.} of the inclination with a characteristic rate:
\begin{align}
\label{tauwave}
\frac{1}{i'}\frac{d i'}{dt} \bigg|_{\rm{d}} &= n' \frac{\xi}{\beta^4} \left( \frac{m'}{M} \right) \left( \frac{\Sigma \, (a')^2}{M} \right)  \equiv \frac{\xi}{\tau_{\rm{wave}}},
\end{align}
where $\xi = 0.544$ is a dimensionless constant \citep{TanakaWard2004,CresswellNelson2008}. 

Of the two effects, consequences of direct inclination damping are easier to understand. It follows from Hamilton's equations that if 
\begin{align}
\label{torquebalance}
\frac{\xi}{\tau_{\rm{wave}}} &\lesssim n' \, s\,  \bigg( \frac{m}{M} \bigg) \bigg( \frac{f_2}{2} \bigg) \simeq n' \, s \, \frac{3}{2} \bigg( \frac{m}{M} \bigg) \bigg( \frac{a}{a'} \bigg)^2,
\end{align}
then disk-driven inclination damping experienced by the outer orbit will result in a gradual decrease of the phase-space area occupied by the orbit, $\mathcal{J}$ \citep{Henrard1993}. In this regime, the dynamical state of the system converges onto an equilibrium point engulfed by the starting trajectory\footnote{In other words, the final evolutionary state is determined by the domain of the phase-space portrait (i.e. inner circulation, resonant, outer circulation) in which the trajectory is initialized.} \citep{BatyginMorbidelli2011}. Simultaneously, the timescale associated with mutual orbit-orbit inclination damping is boosted by the ratio of the inner orbit's angular momentum to that of the outer orbit\footnote{A direct parallel can be drawn between this process and the reason behind why non-zero eccentricity of a close-in orbit can be maintained against tidal dissipation, by secular interactions with a distant companion in the HAT-P-13 system \citep{Batyginetal2009}.}. 

\begin{figure*}[t]
\centering
\includegraphics[width=\textwidth]{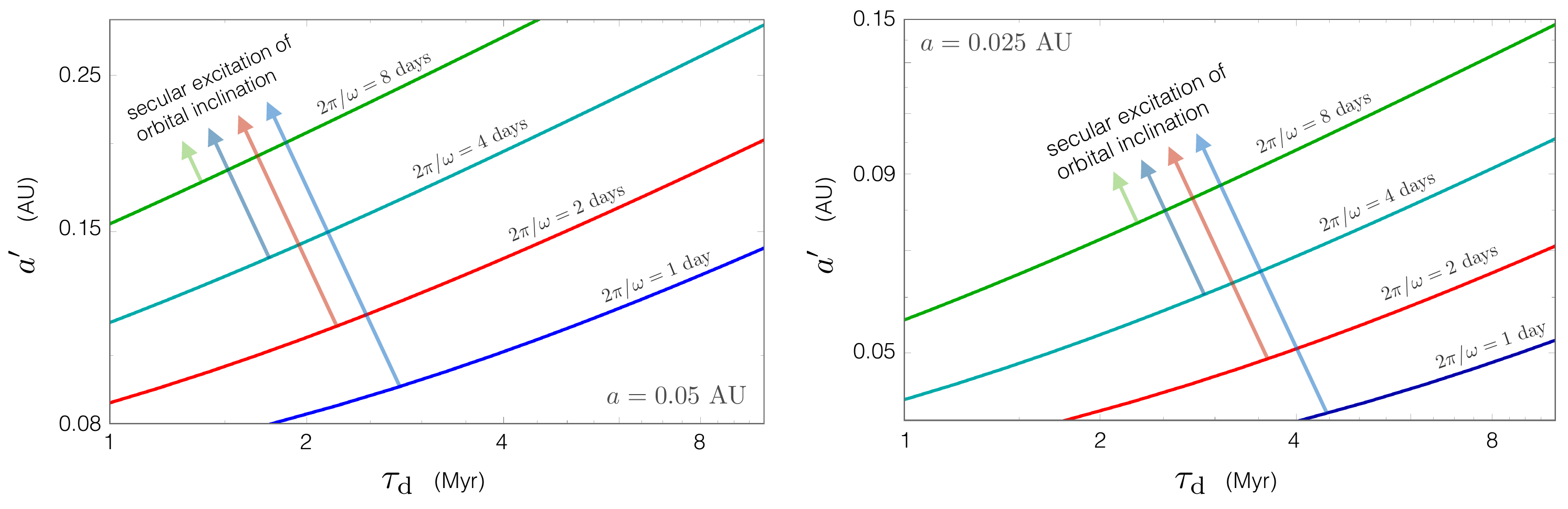}
\caption{Critical test particle semi-major axis beyond which secular resonant excitation of mutual inclination is guaranteed to take place. The shown family of curves spans the observationally relevant range of disk lifetimes, $\tau_{\rm{d}}$, and T-Tauri stellar rotation periods, $2\pi/\omega$, for a hot Jupiter with mass $m=10^{-3} M_{\odot}$ and semi-major axis $a=0.05$ AU (left panel) and $a=0.025$ AU (right panel). The depicted loci approximately follow equation (\ref{crit}), and demonstrate that short disk lifetimes, rapid stellar rotation, and close proximity to the host star are preferred for the operation of the adiabatic excitation process described herein.}
\label{deltaincfig}
\end{figure*}

If on the other hand, the condition (\ref{torquebalance}) is not satisfied, the inclination simply decays to zero. Accordingly, consequences of direct inclination damping can be summarized as follows. During early stages of disk evolution, it establishes initial conditions where all orbits are planar. During late stages of disk evolution, it drives the system to the ``nearest" stable equilibrium. If resonant locking is successful, this reduces the libration amplitude of the harmonic angle. If resonant locking is unsuccessful, inclination decays to zero. While both of these regimes represent distinct possibilities in real protoplanetary disks, it is easy to check that by the time giant planet conglomeration is complete (i.e. disk ages of $\sim 1\,$Myr or more), satisfaction of inequality (\ref{torquebalance}) is likely, given reasonable parameters. 

Consequences of disk-induced nodal regression are somewhat more elusive, and are best illustrated by considering the alteration and evolution of the resonance proximity parameter. Incorporating disk-induced nodal regression (\ref{nodedisk}) into Hamiltonian (\ref{Htp}) and carrying through the analysis as above, we obtain the following expression for the full proximity parameter: 
\begin{align}
\label{deltad}
\delta_{\rm{d}} &= \delta +  \frac{1}{2 \, \beta} \bigg(\frac{64}{f_2^2 \, f_3 \, s^2} \bigg)^{1/3} \bigg( \frac{m_{\rm{d}}}{m} \bigg) \bigg( \frac{a}{a_{\rm{d}}} \bigg),
\end{align}
where $\delta$ is given by equation (\ref{delta}). 

As long as the disk mass is finite, the newly introduced correction causes the true value of the proximity parameter to exceed the ``isolated" quantity given by equation (\ref{delta}). The extent of the correction, however, depends sensitively on the assumed parameters. In particular, provided the variance in T-Tauri stellar rotation periods of $2\pi/\omega \sim 1-10$ days \citep{Herbstetal2007,Afferetal2013} and typical protoplanetary nebula lifetimes of $\tau_{\rm{d}} \sim 1-10$ Myr \citep{Haischetal2001,WilliamsCieza2011}, the disk-induced contribution can be negligible or dominant.

In a regime where the disk correction is negligible (rapid stellar rotation, short disk lifetime), dynamical evolution proceeds as described before and large mutual inclinations can be easily excited in hot Jupiter systems. On the other hand, in a regime where the disk correction plays a commanding role (slow stellar rotation, long disk lifetime), the value of $\delta_{\rm{d}}$ can become exceedingly large early in the nebular phase and approach the asymptotic value given by equation (\ref{deltafin}) from above. In this case, the topology of the Hamiltonian remains consistently similar to that depicted in the right-most panel of Figure (\ref{sfmrfig}) and the trajectory never leaves the immediate vicinity of the equilibrium point within the inner circulation region of the phase-space portrait. Consequently, significant inclination is never excited.

One can easily envision more complex evolutionary sequences for the proximity parameter. For instance, in a limited parameter range, $\delta_{\rm{d}}$ can originate at a large value but drop below zero while the disk is still present and subsequently rise again after nebular dissipation. Such a sequence of events can lead to a ``divergent" encounter with a separatrix, which can pre-excite the mutual inclinations and potentially compromise subsequent establishment of secular resonance\footnote{Whether or not inclination is damped away depends on the remaining disk lifetime and the mass of the exterior planet.} \citep{BorderiesGoldreich1984}. Another alternative is one where $\delta_{\rm{d}}$ rises monotonically, but secular resonance is encountered while the disk is still massive enough to break the resonant lock through non-adiabatic inclination damping (i.e. violation of inequality \ref{torquebalance}).

For the problem at hand, we can translate the bounding criterion to a requirement that \textit{at the epoch of disk dispersal, the proximity parameter is negative.} In doing so, we identify a regime where the post-nebular dynamical state of the planetary system corresponds to the initial condition envisioned in the preceding discussion (represented in Figure \ref{setupfig}), and secular increase of mutual inclination is associated entirely with the fading of the stellar quadrupole moment.

\begin{figure*}[t]
\includegraphics[width=\textwidth]{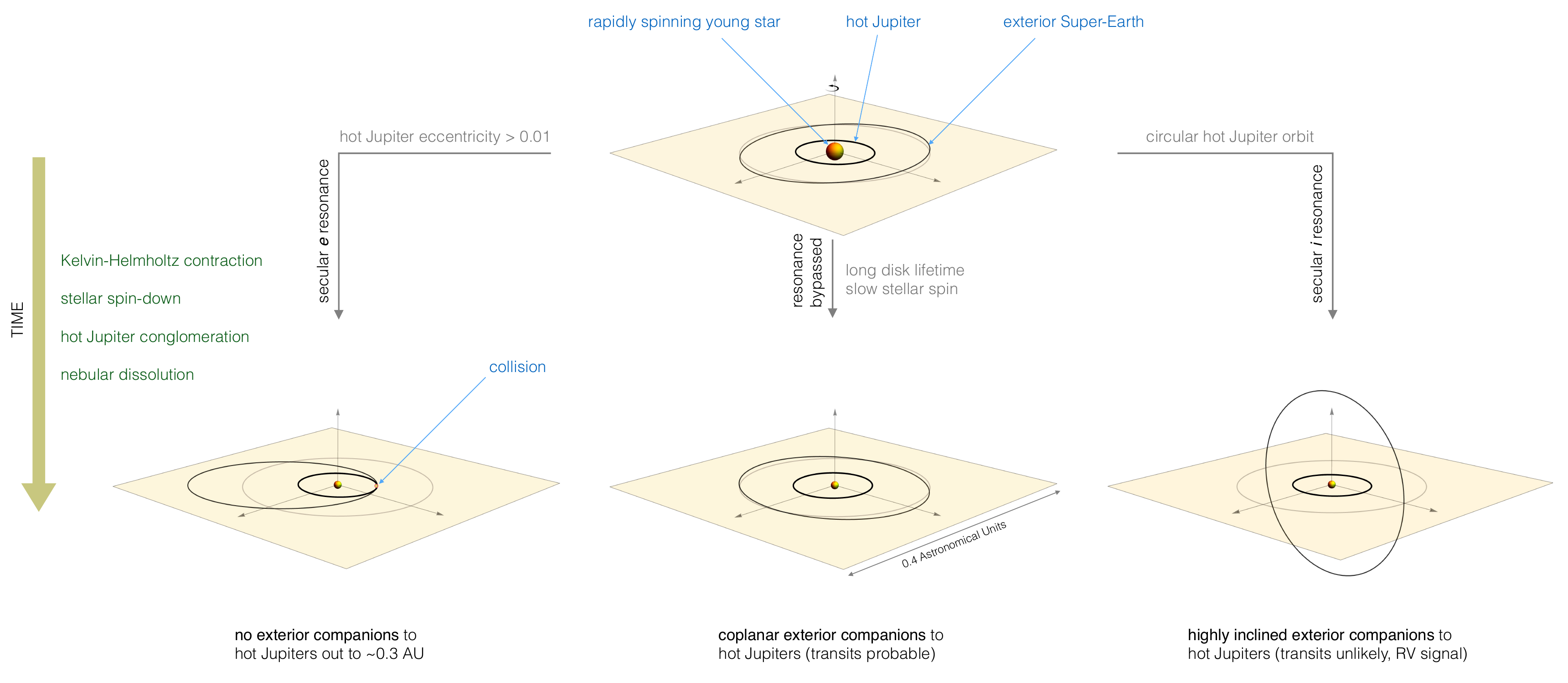}
\caption{Potential outcomes of dynamical evolution of hot Jupiter-hosting planetary systems. An initially nearly-planar, quasi-circular, low-mass multi-planetary system is taken to evolve under the influence of mutual gravitational coupling, interactions with the protoplanetary nebula, as well as the quadrupole field of the young, rapidly rotating star. As the nebula dissipates,  the inner orbit experiences \textit{in situ} conglomeration. Meanwhile, the star undergoes gravitational contraction and loses angular momentum, thereby shedding its quadrupole moment. Cumulatively, these physical processes can give rise to scanning secular resonances that sweep through the inner region of the planetary system. As a result, exterior companions to hot Jupiters can be driven onto intersecting trajectories, or acquire nearly orthogonal orbits, depending on whether hot Jupiters maintain eccentricities above or below $e_{\rm HJ} \sim 0.01$ during the early stages of their lifetimes. On the other hand, coplanarity and dynamical stability can be maintained if disk lifetime is sufficiently long, or stellar rotation is sufficiently slow, to preclude the establishment of secular resonances.}
\label{fig:OutcomesFig}
\end{figure*}

To evaluate the envisioned criterion quantitatively, we must connect the physical evolution of the star to the disk lifetime. During PMS stages of stellar lifetimes, physical evolution is dominated by the release of gravitational energy, and the radius evolution can be expressed approximately as Kelvin-Helmholtz contraction (e.g. \citealt{BatyginAdams2013}):
\begin{align}
\label{Rstar}
R_{\rm{\star}} &\simeq \bigg[ \frac{\G \, M^2}{28 \, \pi \, \sigma \, T_{\rm{eff}}^4\, \tau_{\rm{d}}}  \bigg]^{1/3},
\end{align}
where $\sigma$ is the Stefan-Boltzmann constant, $T_{\rm{eff}}$ is the effective temperature, and as before we have assumed a polytropic equation of state with an index of $3/2$. For a Sun-like star, numerically computed PMS evolutionary tracks are well matched by setting $T_{\rm{eff}} \simeq 4000$ K in the above equation \citep{Siessetal2000}.

Provided a stellar rotation rate at $t=\tau_{\rm{d}}$ (epoch of the onset of photo-evaporation) as well as a semi-major axis and a mass of the hot Jupiter, the above expression can be combined with equation (\ref{delta}) to yield a semi-major axis, $a'$, beyond which secular resonant growth of orbital inclination can be expected to operate. Figure (\ref{deltaincfig}) shows the critical orbital radii for a $m=10^{-3} M_{\odot}$ hot Jupiter with $a=0.025$ AU (right panel) and $a=0.05$ AU (left panel) over the observationally relevant range of stellar spin and disk lifetime parameters. We note further that for well-separated orbits, we can neglect the first term in equation (\ref{delta}) in favor of the second, and replace the Laplace coefficient with its leading order hypergeometric series approximation to derive a simplified form of the criterion: 
\begin{align}
\label{crit}
a' \gtrsim a \, \bigg( \frac{3 \, \mathcal{G} \, m \, a^2}{2 \, k_2 \,\omega^2 \, R_{\star}^5 }   \bigg)^{2/7}.
\end{align}
The nonlinear dependence of the critical semi-major axis, $a'$, on the hot Jupiter semi-major axis, $a$, observed in Figure (\ref{deltaincfig}) is thus qualitatively understood. 

An essentially identical analysis can be carried out for encounters with secular eccentricity resonances, although the evolutionary picture is somewhat more opaque. Compared to the disk-induced nodal regression rate (equation \ref{nodedisk}), the forced apsidal regression rate is diminished by a factor of $\beta$ \citep{Ward1981}. This means that the disk contribution to secular eccentricity evolution is substantially less important, and eccentricity resonances are more readily encountered during the disk-bearing stage of stellar evolution. 

Naively, this would imply that low-mass exterior companions to hot Jupiters are habitually lost to dynamical instabilities (as shown in Figure \ref{eccfig}), followed by collisions with the host stars or hot Jupiters themselves. At the same time, however, eccentricity resonances are more easily overpowered by non-adiabatic damping than their inclination counterparts. In particular, much like inclination damping, eccentricity damping that arises from excitation of spiral density waves by the planet occurs on a timescale $\sim \tau_{\rm{wave}}$ (e.g. \citealt{TanakaWard2004}), while the harmonic term (the equivalent of the RHS of equation \ref{torquebalance}) is smaller by an additional factor of $b_{3/2}^{(2)}/b_{3/2}^{(1)} \simeq a/a'$.

Furthermore, in order for the secular eccentricity resonance to operate in the first place, some mechanism must be invoked to maintain the hot Jupiter eccentricity at non-zero values throughout the disk's lifetime. While certainly not implausible (see e.g. \citealt{GoldreichSari2003,Riceetal2008,DuffellChiang2015}), the extent to which hot Jupiter eccentricities of $e \gtrsim 0.01$ are common during the early stages of planetary system evolution is unknown. In light of the uncertainty inherent to the eccentricity counterpart of the presented calculation, detailed dynamical simulations are required to address its importance. Although comprehensive numerical experiments of this sort are beyond the scope of this work, they pose an important direction for the development of the presented theory.

\section{Discussion}

In this work, we have considered the possibility that the origins of the hot Jupiter population of extrasolar planets, long thought to originate from long-range radial migration, can be successfully explained within the framework of \textit{in situ} formation. We first used a planetary evolution code \citep{Hubickyj2005} to demonstrate that rapid accretion of gas onto protoplanetary cores under nebular conditions appropriate to the inner-most regions of protoplanetary nebulae, can be readily triggered provided a core whose mass exceeds $M_{\rm core} \gtrsim 15\, M_{\oplus}$. \textit{In situ} formation of hot Jupiters thus implies that the largest sub-critical cores will have masses no bigger than $M_{\rm p}\sim30\,M_{\oplus}$, in agreement with the observed distribution of planets shown in Figure (\ref{fig:MRfig}). 

If \textit{in situ} formation represents the dominant channel for hot Jupiter generation, then the progenitor population of close-in giant planets are Super-Earths that occupy a similar orbital range. In light of the prevalence of planetary multiplicity among Super-Earth systems, it is reasonable to presume that giant planet conglomeration frequently occurs in presence of low-mass companion planets that reside on exterior orbits. Such a scenario entails an accompanying dynamical evolution that in turn yields readily testable observational consequences. 

The process of planetary conglomeration paired with Kelvin-Helmholtz contraction, and eventual spin-down of the star, drives scanning secular resonances that act to alter the dynamical state of exterior Super-Earths. By employing both an analytic treatment and $N$-body simulations, we find that Super-Earths with orbital radii $a\lesssim0.3\,$AU can be coerced onto crossing orbits and consumed by the hot Jupiter, if its orbital eccentricity exceeds $e_{\rm HJ}\gtrsim 0.01$. In systems that are so destabilized, we expect to observe a remnant, dynamically undisturbed population of planets with orbital periods well beyond $P>50\,$days. Additionally, we expect to find evidence of metallicity enrichment within the hot Jupiters themselves. That is, the \textit{in situ} formation scenario predicts that hot Jupiters without Super-Earth companions should, statistically, have larger core masses.

The Saturn-mass planet HD 149026\,b \citep{Satoetal2005}, as well as similar high-core mass planets such as HAT-P-25\,b and Kepler 424\,b, present possible archetypal examples of destabilized systems in which one or more Super-Earths were accreted to grow the hot Jupiter's core to the large inferred size. Indeed, for a planet such as HD 149026\,b, \textit{in situ} formation is strongly favored, and may in fact be required, to achieve $M_{\rm core}\sim 100M_{\oplus}$. At HD 149026\,b's current orbital radius ($a\simeq0.04\,$AU) the ratio of the planet's surface escape speed to its orbital velocity, $v_{\rm esc}/v_{\rm Kep}\sim 0.2$ is quite small. As a result, the planet cannot readily eject smaller companion planets from the system when close encounters occur, and over time, the hot Jupiter is much more likely to accrete any erstwhile low-mass neighbors that evolve onto intersecting orbits. By contrast, rather specific assumptions regarding disk structure and evolution would likely have to be made to explain the origins of this object within the context of migration theory.

Dramatically different final outcomes are foreseen for companion low-mass planets in situations where the hot Jupiter's orbit remains nearly circular throughout the disk-bearing stage of stellar lifetime.  In this case, we find that a commensurability is achieved between the nodal regression rates of the hot Jupiter and any Super-Earths on exterior ($a\gtrsim 0.1 - 0.2\,$AU) orbits. As the star's rotation slows, its quadrupole moment nearly disappears, but the regression rate commensurability between the hot Jupiter and the Super-Earth companion is maintained (that is, the system finds itself locked in a stable secular inclination resonance). Consequently, the Super-Earth is driven to a highly inclined orbit that is nearly orthogonal to the orbital plane of the hot Jupiter. This unexpected result appears to be robust for a range of plausible initial configurations, and therefore provides a strong, readily falsifiable prediction. In particular, short disk lifetimes as well as rapid T-Tauri stellar rotation are preferred for generation of orthogonal orbits.

Ancillary effects associated with the presence of the protoplanetary disk during the early epochs of planetary system evolution affect the establishment of secular resonances on a detailed level. Specifically, the aforementioned process of adiabatic inclination growth fails to operate for sufficiently compact orbits, due to the perturbing potential of the protostellar disk (a simple criterion for the onset of resonance is given by equation \ref{crit}). Accordingly, the \textit{in situ} formation paradigm also points towards the existence of a population of planar, circular companions to hot Jupiters. The three possible outcomes of dynamical evolution ensuing \textit{in situ} formation of hot Jupiters are summarized in Figure (\ref{fig:OutcomesFig}).

The recently discovered companions to the solar-type star WASP-47 may represent an undisrupted outcome of the \textit{in situ} formation process. WASP-47\,b ($M_{\rm P}\sim1M_{\rm Jup}$, $P=4.159\,$d) is a highly representative hot Jupiter that was originally detected from the ground \citep{Hellier2012}. Follow-up observations using the \textit{Kepler} space telescope in its K2 operation mode have led to the discovery of two additional transiting Super-Earth planets in the system \citep{Beckeretal2015}; an inner, ultra-short period planet, ``c'', with $M_{\rm P}<9 M_{\rm \oplus}$ and $P=0.790\,$days, and an outer companion, ``d'', with $M_{\rm P}\sim 9 M_{\rm \oplus}$ and $P=9.031\,$days. 

Our \textit{in situ} formation scenario implies that the bulk of WASP-47\,b's gas was accreted while the planet maintained a near-circular orbit, and that the accompanying Super-Earths (planets c and d) were thus able to avoid orbital instabilities. Assuming a representative disk lifetime of $\sim 3\,$Myr and a T-Tauri stellar rotation period of $\sim 3\,$days, our dynamical calculations imply that the observed orbits could have avoided the excitation of mutual inclination due to the presence of the disk. However, any putative low-mass objects residing on substantially more distant orbits could have been forced onto highly inclined trajectories.

Because the core accretion process exhibits a very weak dependence on orbital radius, giant planet formation is expected to occur over the entire orbital range occupied by the Super-Earths \citep{LeeChiang2014}. Accordingly, the picture presented herein places hot and warm Jupiters into essentially the same evolutionary context. However, due to the fact that secular resonance only affects low mass objects that reside on exterior orbits, interior companions to warm Jupiters are expected to remain in unperturbed, coplanar orbits within the framework of our dynamical model. Remarkably, the recent analysis of \citet{Huang2016} reveals that approximately half of all warm Jupiters in the \textit{Kepler} sample are accompanied by low-mass planets, with a strong preference for shorter orbital periods. Indeed, this finding signals consistency of the observational data with \textit{in-situ} formation of close-in giant planets.

The process of \textit{in situ} close-in giant planet formation, followed by dynamical evolution facilitated by changes in the physical character of the system, has a number of implications for existing hot Jupiter observations. Within the context of our model, commonly observed spin-orbit misalignments are most readily ascribed to primordial star-disk inclinations \citep{Bateetal2010,Lai2011,Batygin2012}, while the associated correlation between orbital obliquity and stellar effective temperature can be attributed to magnetic disk-star coupling during the T-Tauri stages of host stellar lifetimes \citep{SpaldingBatygin2015}. Additionally, dynamical excitation of orbit-orbit inclinations provides a natural resolution to the scarcity of hot Jupiter systems with multi-transiting companions. The hypothesized existence of a population of low-mass planets beyond the orbits of known hot Jupiters is readily testable with existing radial velocity spectrographs such as Keck, HARPS, or APF. In short, our theoretical framework will be subject to near-immediate observational tests.
\\
\\
\textbf{Acknowledgments}  \\ 
We are thankful to Dave Stevenson, Chris Spalding, Mike Brown and Heather Knutson for inspirational conversations, as well as to Eugene Chiang for providing a thorough review of the paper, which led to a substantial improvement of the manuscript.

\end{document}